\documentclass[twoside,twocolumn,9pt]{article}
\usepackage{extsizes}
\usepackage[super,sort&compress,comma]{natbib} 
\usepackage[version=3]{mhchem}
\usepackage[left=1.5cm, right=1.5cm, top=1.785cm, bottom=2.0cm]{geometry}
\usepackage{balance}
\usepackage{times,mathptmx}
\usepackage{sectsty}
\usepackage{graphicx} 
\usepackage{multirow}
\usepackage{lastpage}
\usepackage[format=plain,justification=justified,singlelinecheck=false,font={stretch=1.125,small,sf},labelfont=bf,labelsep=space]{caption}
\usepackage{float}
\usepackage{fancyhdr}
\usepackage{pdflscape} 
\usepackage{tabularx}
\usepackage{fnpos}
\usepackage[english]{babel}
\addto{\captionsenglish}{%
  
}
\usepackage{array}
\usepackage{droidsans}
\usepackage{charter}
\usepackage[T1]{fontenc}
\usepackage[usenames,dvipsnames]{xcolor}
\usepackage{setspace}
\usepackage[compact]{titlesec}
\usepackage{hyperref}

\usepackage{epstopdf}

\definecolor{cream}{RGB}{222,217,201}

\begin{document}

\makeFNbottom
\makeatletter
\renewcommand\LARGE{\@setfontsize\LARGE{15pt}{17}}
\renewcommand\Large{\@setfontsize\Large{12pt}{14}}
\renewcommand\large{\@setfontsize\large{10pt}{12}}
\renewcommand\footnotesize{\@setfontsize\footnotesize{7pt}{10}}
\makeatother

\renewcommand{\thefootnote}{\fnsymbol{footnote}}
\renewcommand\footnoterule{\vspace*{1pt}%
\color{cream}\hrule width 3.5in height 0.4pt \color{black}\vspace*{5pt}} 
\setcounter{secnumdepth}{5}

\makeatletter 
\renewcommand\@biblabel[1]{#1}            
\renewcommand\@makefntext[1]%
{\noindent\makebox[0pt][r]{\@thefnmark\,}#1}
\makeatother 
\renewcommand{\figurename}{\small{Fig.}~}
\sectionfont{\sffamily\Large}
\subsectionfont{\normalsize}
\subsubsectionfont{\bf}
\setstretch{1.125} 
\setlength{\skip\footins}{0.8cm}
\setlength{\footnotesep}{0.25cm}
\setlength{\jot}{10pt}
\titlespacing*{\section}{0pt}{4pt}{4pt}
\titlespacing*{\subsection}{0pt}{15pt}{1pt}

\fancyfoot{}
\fancyfoot[LO,RE]{\vspace{-7.1pt}\includegraphics[height=9pt]{head_foot/LF}}
\fancyfoot[CO]{\vspace{-7.1pt}\hspace{13.2cm}\includegraphics{head_foot/RF}}
\fancyfoot[CE]{\vspace{-7.2pt}\hspace{-14.2cm}\includegraphics{head_foot/RF}}
\fancyfoot[RO]{\footnotesize{\sffamily{1--\pageref{LastPage} ~\textbar  \hspace{2pt}\thepage}}}
\fancyfoot[LE]{\footnotesize{\sffamily{\thepage~\textbar\hspace{3.45cm} 1--\pageref{LastPage}}}}
\fancyhead{}
\renewcommand{\headrulewidth}{0pt} 
\renewcommand{\footrulewidth}{0pt}
\setlength{\arrayrulewidth}{1pt}
\setlength{\columnsep}{6.5mm}
\setlength\bibsep{1pt}

\makeatletter 
\newlength{\figrulesep} 
\setlength{\figrulesep}{0.5\textfloatsep} 

\newcommand{\topfigrule}{\vspace*{-1pt}%
\noindent{\color{cream}\rule[-\figrulesep]{\columnwidth}{1.5pt}} }

\newcommand{\botfigrule}{\vspace*{-2pt}%
\noindent{\color{cream}\rule[\figrulesep]{\columnwidth}{1.5pt}} }

\newcommand{\dblfigrule}{\vspace*{-1pt}%
\noindent{\color{cream}\rule[-\figrulesep]{\textwidth}{1.5pt}} }

\makeatother

\twocolumn[
  \begin{@twocolumnfalse}
\vspace{3cm}
\sffamily

 \noindent\LARGE{\textbf{Crystal structure, Chemical Bonding, Electrical and Thermal Transport in Sc$_5$Rh$_6$Sn$_{18}$}} \\

  \noindent\large{Manuel Feig,\textit{$^{a,b}$} Lev Akselrud,\textit{$^{b,c}$} Walter Schnelle,\textit{$^{b}$}  Vadim Dyadkin,\textit{$^{d}$} Dmitry Chernyshov,\textit{$^{d}$} Alim Ormeci,\textit{$^{a,b}$} Paul Simon,\textit{$^{b}$} Andreas Leithe-Jasper,\textit{$^{b}$} and Roman Gumeniuk$^{\ast}$\textit{$^{a}$} } \\

 \noindent\normalsize{Single crystals of Sc$_5$Rh$_6$Sn$_{18}$ were grown from Sn-flux. The crystal structure (SG: $I4_1/acd$, \textit{a} = 13.5529(2) \AA, \textit{c} =  27.0976(7) \AA) was studied by high-resolution X-ray diffraction on powder and single crystal material as well as by TEM. All methods confirm it to crystallize with Sc$_5$Ir$_6$Sn$_{18}$ (space group $I4_1/acd$) type of structure. The performed structural studies suggest also the presence of local domains with broken average translational symmetry. Analysis of the chemical bonding situation reveal highly polar Sc2-Sn1, Sn-Rh and Sc2-Rh bonds, two- and three-center bonds involving Sn-atoms as well as ionic nature of Sc1 bonding. The thermopower of Sc$_5$Rh$_6$Sn$_{18}$ is isotropic, small and negative (i.e. dominance of electron-like charge carriers). Due to structural disorder, the thermal conductivity is lowered in comparison with regular metallic systems.} \\

 \end{@twocolumnfalse} \vspace{0.6cm}

  ]

\renewcommand*\rmdefault{bch}\normalfont\upshape
\rmfamily
\section*{}
\vspace{-1cm}


\footnotetext{\textit{$^{a}$~Institut f\"ur Experimentelle Physik, TU Bergakademie
		Freiberg, Leipziger Stra{\ss}e 23, 09596 Freiberg, Germany; E-mail: roman.gumeniuk@physik.tu-freiberg.de}}
\footnotetext{\textit{$^b$~Max-Planck-Institut f\"ur Chemische Physik fester Stoffe,
		N\"othnitzer Stra{\ss}e 40, 01187 Dresden, Germany}}
\footnotetext{\textit{$^{c}$~Ivan Franko National University of Lviv, Kyryla and Mefodiya Str. 6, UA-79005, Lviv, Ukraine}}
\footnotetext{\textit{$^{d}$~Swiss-Norwegian Beamlines at ESRF, CS 40220, 38043 Grenoble Cedex 9, France}}





\section{Introduction}

After their discovery in the early 1980\textit{s} by Remeika et al.\cite{Remeika1980}, a new series of primitive and face-centred cubic and tetragonal Sn-rich compounds containing rare-earth, alkaline-earth, actinide as well as transition 3$d$- and 4$d$-metals became immediately an object of numerous studies\cite{gum2018Rem}. Continuous interest in these phases is mainly due to the exciting superconducting properties and easy synthesis of nicely shaped large single crystals (i.e. by flux methods).

However, as it turned out in recent investigations, the determination of both the crystal structure and the chemical composition of such stannides is challenging. They reveal both concentration- and temperature-induced phase transitions. For example, La$_3$Rh$_4$Sn$_{13}$ crystallizes with space group (SG) $I4_132$, $a \approx 19$ \AA \cite{Bordet1991}, while La$_{3+x}$Rh$_4$Sn$_{13-x}$ ($x = 1$) is primitive cubic (SG $Pm\overline{3}m$, $a \approx 9$ \AA). \cite{Anand2011} On the other hand, body centred cubic (SG $I4_132$) Ce$_3$Rh$_4$Sn$_{13}$ is observed for $T < 350$\, K, while the primitive cubic modification (SG $Pm\overline{3}m$) exists at higher temperatures.\cite{kuo2018}  Also, stannides with approximate composition $\sim$$M$\{Co, Rh\}$_{1.2}$Sn$_4$ ($M$ = Sc, Y, heavier rare-earth metal), which have initially been reported to be primitive tetragonal with unit cell parameters $a \approx 13$ \AA\, and $c \approx 9$ \AA\cite{Remeika1980,Espinosa1980,Cooper1980,Espinosa1982}, were later shown to possess composition 5:6:18 and to be body-centred tetragonal (SG $I$4$_1$/$acd$) with tripled lattice parameter $c$ (i.e.  $a$\,$\approx$\,13\, \AA\, and $c$\,$\approx$\,27\,\AA).\cite{Chenavas1982}$^{,}$\cite{Miraglia1987} The first refinement of such a structural model was performed for Tb$_5$Rh$_6$Sn$_{18}$\cite{Miraglia1987} and revealed a statistical mixture of Sn and Tb to occupy one of the crystallographic sites [i.e. the (Sn$_{1-x}$Tb$_x$)Tb$_4$Rh$_6$Sn$_{18}$ composition]. Later studies showed this structure type to be characterized either by unphysical anisotropies of thermal displacements, as in the case of Sc$_5$Co$_6$Sn$_{18}$\cite{Kotur1999, Lei2009} or by numerous splits of crystallographic positions occupied by Sn-atoms (Sc$_5$Ir$_6$Sn$_{18}$\cite{Levy2019}). The crystal structure of Sc$_5$Rh$_6$Sn$_{18}$ has never been refined up to now. The unit cell parameters $a$\,$\approx$\,13.5\, \AA\, and $c$\,$\approx$\,27.1\,\AA\, and SG $I$4$_1$/$acd$ were reported in \cite{Chenavas1982} and authors of \cite{Kase2012}$^,$\cite{Bhattacha2018} assumed it to crystallize with the Tb$_5$Rh$_6$Sn$_{18}$ structure type.

Being a superconductor with $T_\mathrm{c}$ $\approx$~5~K and relatively high critical magnetic field ($B_\mathrm{c2}$ $\approx$~8~T), Sc$_5$Rh$_6$Sn$_{18}$ became an object of numerous investigations.\cite{Remeika1980,Espinosa1980,Espinosa1982,Venturini1986,Kase2012} A very recent muon-spin-rotation ($\mu$SR) study \cite{Bhattacha2018} performed on crystals of this stannide indicated a time-reversal symmetry (TRS) breaking of the superconducting state and thus, unconventional superconductivity with either a singlet d + id state with a line node or, alternatively, non unitary triplet pairing with point nodes \cite{Bhattacha2018}. This finding raises the question about the inversion center in the crystal structure of Sc$_5$Rh$_6$Sn$_{18}$. As it is well known, non-centrosymmetric superconductors (SC) are preferable candidates for unconventional superconducting properties, which includes TRS breaking as well.\cite{Bauer2012}

Taking all these observations into account as well as a strong structural disorder reported for isostructural Sc$_5$Ir$_6$Sn$_{18}$ superconductor \cite{Levy2019} we performed an extensive study of the crystal structure of Sc$_5$Rh$_6$Sn$_{18}$ applying temperature dependent high-resolution powder and single crystal X-ray diffraction as well as transmission electron microscopy. To understand the obtained structural model the analysis of the chemical bonding situation assuming different structural models is performed. These investigations and simulations revealed Sc$_5$Rh$_6$Sn$_{18}$ to crystallize with the centrosymmetric Sc$_5$Ir$_6$Sn$_{18}$ structure type. Also, we demonstrate the consequences of the structural disorder of the compound for its electrical and thermal transport properties.

\section{Experimental}

\subsection{Synthesis}

\begin{figure}
	\centering
	\includegraphics[width=6cm, height=6cm]{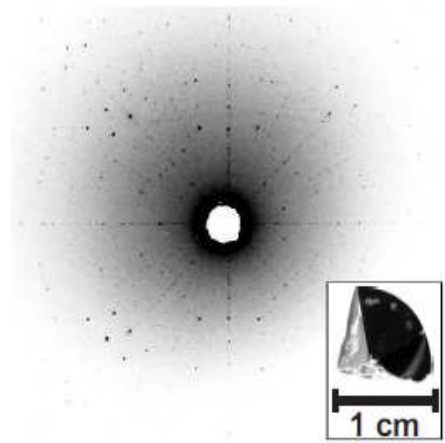}
	\caption{Typical Laue pattern of a Sc$_5$Rh$_6$Sn$_{18}$ single crystal (shown in inset).}
	\label{fig:cry}
\end{figure}

Samples with nominal composition Sc$_5$Rh$_6$Sn$_{18}$ and total mass of 2\,g were prepared from scandium ingots (Dr.\ Lamprecht, 99.5\,wt.\,\%), rhodium ingots (Chempur, 99.95\,wt.\,\%) and tin (Chempur 99.9999\,wt.\,\%) by arc melting (mass losses $<0.1$\,\%). Obtained buttons with 10 g of tin excess were placed in glassy carbon crucibles, sealed in a tantalum tube and enclosed in an evacuated silica ampoule. All manipulations were performed inside an argon-filled glove box [$p$(O$_2$,H$_2$O) $\leq$ 1 ppm]. To grow the crystals the samples were heated up to 1473\,K within 24\,h, held at this temperature for 48\,h and then cooled down to 673\,K within 1100\,h. The tin excess was removed from the samples by centrifugation at 873\,K. Obtained buttons were additionally washed in diluted (5\,\%) HCl acid. A typical Sc$_5$Rh$_6$Sn$_{18}$ crystal of a pyramidal shape with $\sim 1 \times 1 \times 1$ cm dimensions together with a well resolved Laue pattern, are shown in Fig.\ \ref{fig:cry}. Further, the crystals were oriented and bars parallel to $c$- and $ab$-directions were cut for the physical properties measurements.

\subsection{Differential thermal analysis}

Differential thermal analysis (DTA) was performed on a piece of single-crystal using a \textit{Netzsch DSC 404C}, under a steady argon flow at a heating rate of 5\,K\,min$^{-1}$. The melting point of Sc$_5$Rh$_6$Sn$_{18}$ was determined via the peak onset to be 1275\,K.

\subsection{X-ray diffraction}

Laboratory X-ray diffraction (XRD) of the powdered crystals was performed on a HUBER G670 imaging plate Guinier camera (Cu$K_{\alpha1}$ radiation, $\lambda$\,=\,1.54056 \r A). The \emph{WinXpow} program package  was used for phase analysis \cite{STOE}. 

Temperature-dependent synchrotron XRD was performed on powder at the Rossendorf beamline BM20 of the ESRF using synchrotron radiation of wavelength $\lambda$ = 0.45932\,\r A. Sieved, powdered samples (fraction \textless\,20\,$\mu$m) were measured in quartz glass capillaries in the temperature range from 100\,K to 300\,K using a nitrogen cryostream apparatus to cool the samples. Diffraction patterns were recorded from 5$^{\circ}$ to 26$^{\circ}$ 2$\theta$ with $\Delta$2$\theta$ = 0.002$^{\circ}$ steps and a counting time of 0.5\,s per step.

Single crystal XRD data were also collected with a Rigaku AFC7 diffractometer (Mo$K_{\alpha}$ radiation, $\lambda$\,=\,0.71073 \r A) equipped with Saturn724+ CCD detector. 

Single crystal XRD data were collected in the temperature range 100\,K--300\,K (cooling with a nitrogen cryostream apparatus) at the BM01 beamline of the European Synchrotron Radiation Facility (ESRF, Grenoble) with $\lambda$ = 0.73331\,\AA. The images acquired on a Pilatus2M detector \cite{Eikenberry2003} were pre-processed with the \textit{SNBL toolbox} \cite{Dyadkin2016} and subsequently processed using the \textit{CrysAlis} \cite{Crysalis} software. 

The lattice parameter refinement by least-squares fitting, Rietveld refinement as well as refinements of single crystal data have been done using the \emph{WinCSD} program package  \cite{Akselrud2014}.

\subsection{Metallography}

For metallographic analysis the synthesized crystals were embedded in conductive resin and then ground, polished and finished with diamond abrasives. The metallographic microstructure was characterized by light-optical microscopy (Zeiss Axioplan 2) and by energy-dispersive X-ray spectroscopy (EDXS) on a JEOL JSM 6610 scanning electron microscope equipped with an UltraDry EDS detector. No additional phases were observed on the surface of the crystals and their composition on the basis of 8 independent measurements was found to be Sc$_{5.0(1)}$Rh$_{6.1(1)}$Sn$_{18.2(2)}$.

\subsection{Transmission Electron Microscopy}

For transmission electron microscopy (TEM) the powdered sample was dispersed in methanol. Several drops of the dispersion were loaded on a carbon coated copper grid and transferred to the microscope after complete dryness. The Quantifoil S7/2 (100-mesh hexagonal) copper grids were covered with 2 nm carbon film (Quantifoil Micro Tools, Jena, Germany). High-resolution TEM (HRTEM) imaging of the sample was performed on a FEI Tecnai F30 with a field-emission gun at an acceleration voltage of 300\,kV. The point resolution amounted to 1.9\,\r A, and the information
limit amounted to about 1.2\,\r A. The microscope was equipped with a wide-angle slow-scan CCD camera (MultiScan, 2k$\times$2k pixels; Gatan Inc., Pleasanton, CA, USA). The analysis of the TEM images was made with the \textit{Digital Micrograph} software (Gatan, USA).

\subsection{Electrical and thermal transport properties}

The Seebeck coefficient of the thermopower, the thermal conductivity as well as the resistivity were measured in a commercial system (PPMS, Quantum Design) on the TTO option.

\subsection{Theoretical calculations}

The electronic structure calculations were carried out within the local density approximation (LDA) to the density functional theory (DFT) using the all electron full-potential local orbital method (FPLO) \cite{Koepernik1999FPLO}. The Perdew-Wang parametrization for the exchange-correlation functional was applied \cite{perdew1992}. Sn 4$s$, 4$p$, 4$d$, Rh 4$s$, 4$p$ and Sc 3$s$, 3$p$ states were treated as semicore. The first Brillouin zone (BZ) was sampled with a 6\,$\times$6\,$\times$6 mesh and the linear tetrahedron method was employed for BZ integrations. 

The position space chemical bonding analysis was based on the topological analysis of the electron density (ED) and electron localizability indicator (ELI) \cite{Kohout2004}. The ELI-D representation of  ELI was used \cite{Kohout2006,Kohout2007}. Both ED and ELI-D were computed by a module implemented in the FPLO method \cite{Ormeci2006}, and their topological analysis was performed by the program DGrid \cite{DGrid}. The underlying theory of this approach is provided by Bader's quantum theory of atoms in molecules (QTAIM) \cite{Quantumtheory}. The basin intersection technique of Raub and Jansen was employed to determine the atoms participating in the bonds \cite{Raub2001}.

\section{Results and discussion}

\subsection{Crystal structure}

\subsubsection{Single crystal and powder X-ray diffraction}

All peaks collected during the room temperature single crystal diffraction measurement on Sc$_5$Rh$_6$Sn$_{18}$ at the BM01 beamline could be indexed with the unit cell parameters given in Table\ \ref{tbl:struct_rt}. No satellite reflections which could indicate either modulation or changes in unit cell dimensions were observed. 

The analysis of the extinction conditions indicated the solely possible SG $I4_1/acd$. The positions of all atoms were successfully localized by the direct methods. They were the same as those reported for Tb$_5$Rh$_6$Sn$_{18}$ type.\cite{Miraglia1987} However, the refinement of atomic coordinates and isotropic displacement parameters with extinction correction converged with rather high reliability factors $R_F$ = 0.110; $R_W$ = 0.118 and residual electronic peaks of -8.1/10.4 e \r A\,$^{-3}$. Also, $B_\mathrm{iso}$ = 3.3(1) for the Sc1-atom was by a factor of $\sim$3 larger than that for all other atoms. By setting the occupation parameters of Sc1 to 0.92(1) and refining the anisotropic displacement parameters for all atoms (Table\ \ref{tbl:powder-nosplit} and \ref{tbl:bvalues}) we were able to obtain both low reliability factors and residual electronic density (Table\ \ref{tbl:struct_rt}). However, a closer look at the $B_\mathrm{anis}$ values (Table\ \ref{tbl:bvalues}) reveals a pronounced anisotropy for almost all Sn-atoms (e.g. $B_{33}\,\approx\,5B_{11}$ for Sn2; $B_{33}\,\approx\,6B_{11}$ for Sn3; $B_{22}\,\approx\,4B_{11}$ for Sn5 $etc$, see Fig. \ref{fig:Anis}b).

\begin{figure}[htb]
	\includegraphics[width=\linewidth]{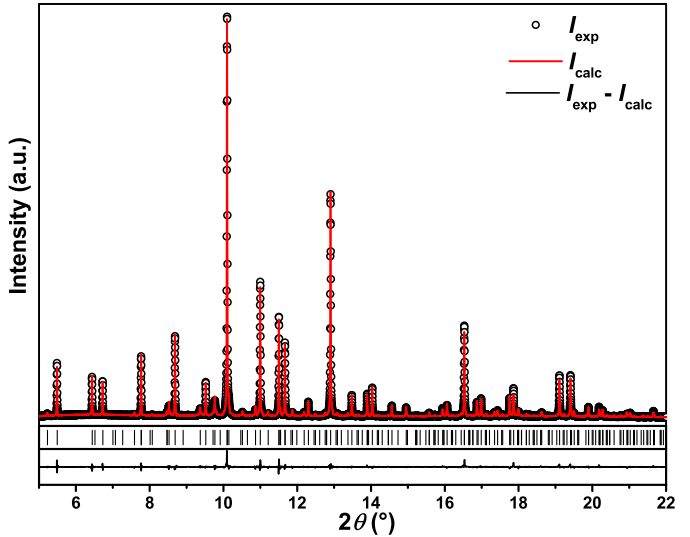}
	\caption{(Color online) {Powder XRD pattern for Sc$_5$Rh$_6$Sn$_{18}$} at room temperature.}
	\label{fig:diff}
\end{figure}

\begin{figure}
	\includegraphics[width=\linewidth]{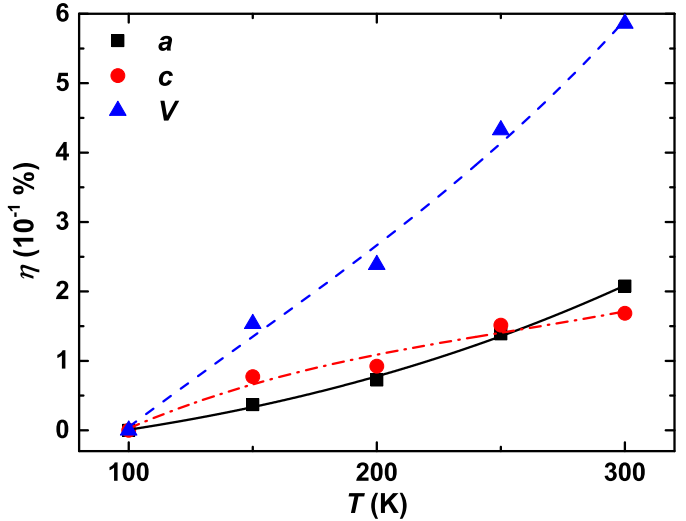}
	\caption{Relative thermal expansion $\eta$ for Sc$_5$Rh$_6$Sn$_{18}$. The lines are guides for the eye.}
	\label{fig:latticeparam}
\end{figure}

\begin{figure*}
	\includegraphics[width=15cm, height=15cm]{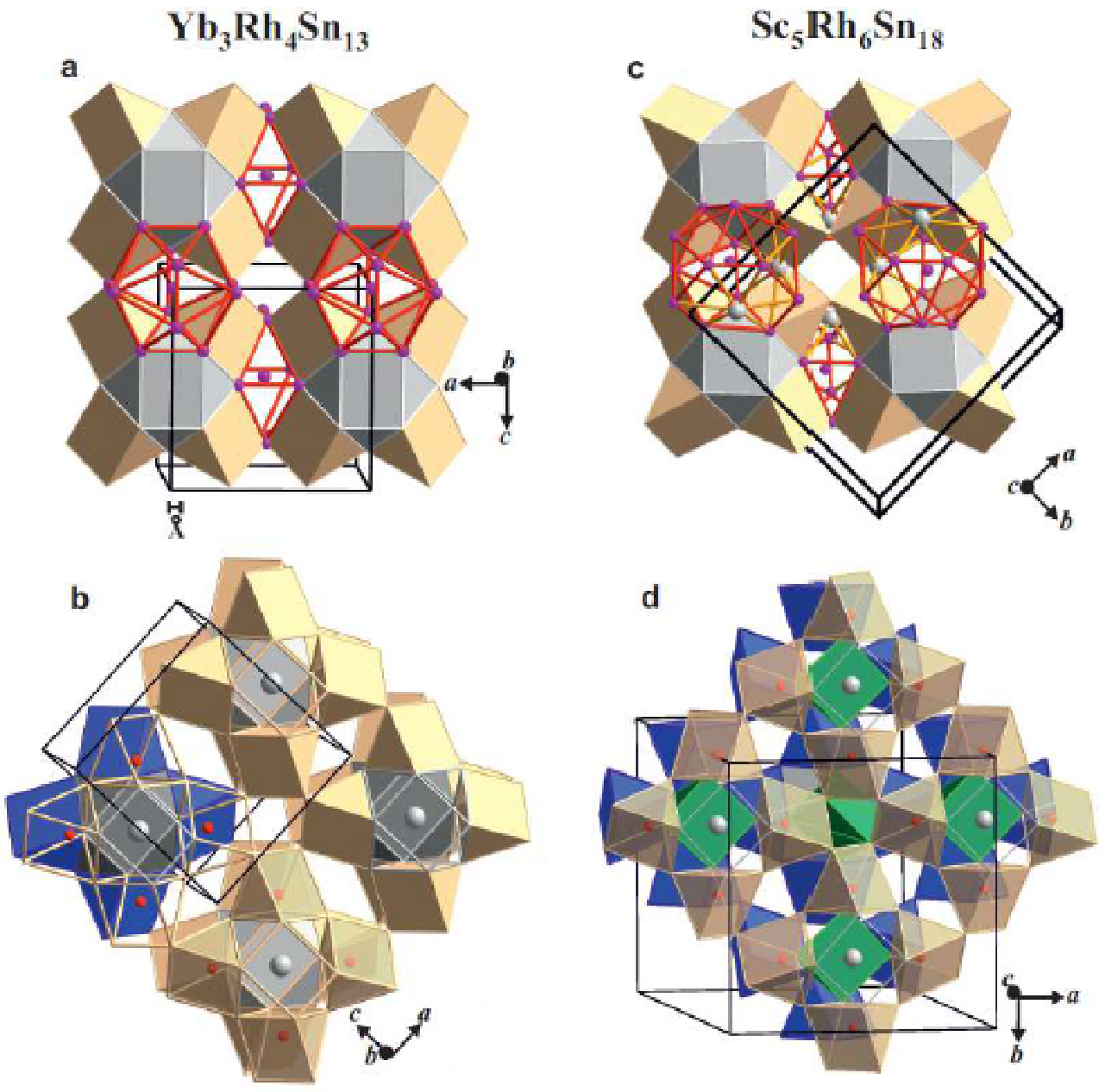}
	\centering
	\caption{Corner sharing trigonal prisms arrays (tan) together with [\textit{M}Sn$_{12}$] cuboctahedra (grey, \textit{M} = Yb, Sc1) and icosahedra (a) or [Sn1Sc2Sn$_{15}$] polyhedra (c) in the structures of Yb$_3$Rh$_4$Sn$_{13}$ type and idealized (Tb$_5$Rh$_6$Sn$_{18}$ type\cite{Miraglia1987}) Sc$_5$Rh$_6$Sn$_{18}$, respectively. b) Condensed columns of cuboctahedra (transparent or grey) extending in $b$-direction with differently oriented trigonal prisms (tan or blue) in the structure of Yb$_3$Rh$_4$Sn$_{13}$ d) Separated cuboctahedra (transparent or grey) as well as 3 layers (tan, green and blue) of corner sharing trigonal prisms in Sc$_5$Rh$_6$Sn$_{18}$. (Sc - grey balls, Rh - red balls, Sn - pink balls).}
	\label{fig:str}
\end{figure*}

Taking into account that the anisotropy is not systematic (which indicates the absorption correction to be performed properly) we assumed it to originate from mechanical stresses and strains appearing in the small crystals while crushing the initial sample. Therefore, in a further step prior to the single crystal diffraction measurement, stress annealing of the small crystals at 1070\, K for 2h was performed. However, further crystal structure refinement resulted in the same partial occupancy by Sc1 of its crystallographic site and unacceptable anisotropic displacement parameters (Table\ \ref{tbl:struct_rt}, \ref{tbl:powder-nosplit} and \ref{tbl:bvalues}). Similar effects were reported for refined structures of isostructural compounds Sc$_5$Co$_6$Sn$_{18}$ \cite{Kotur1999,Lei2009} and Sc$_5$Ir$_6$Sn$_{18}$\cite{Levy2019}.

Interestingly, refinement of the crystal structure of Sc$_5$Rh$_6$Sn$_{18}$ from high resolution synchrotron powder XRD (Table\ \ref{tbl:struct_rt}) (experimentally measured, theoretically calculated and differential profiles are given in Fig.\ \ref{fig:diff}) resulted in partial occupancies by Sn1, Sn4, Sn5 and Sn6 of their crystallographic positions (Table\ \ref{tbl:powder-nosplit}) (if one would set the occupational parameter $G = 1$ for mentioned atoms, this would lead to the increase of the $R_I$ by $\sim$2\, \%).

These observations prompted us to look for an alternative structural model. Taking the model with enhanced reliability factors and residual electronic density (see above) we performed differential Fourier syntheses to localize possible further atomic positions. However, the largest electronic density peaks were localized in close vicinity to Sn2 and Sn3 positions, which indicated them to be split. Other small peaks detected near Sn4-, Sn5- and Sn7-atoms could be implemented into the structural model by shifting these atoms off the centers (i.e. by doubling of the multiplicity of the corresponding Wyckoff sites and thus, $G\,\approx0.5$) (Table\ \ref{tbl:crys3}, \ref{tbl:crys4}). The comparison of the obtained model with that of the initial Tb$_5$Rh$_6$Sn$_{18}$ type is presented in Fig.\ \ref{fig:Baernighausen}.

To prove the correctness of such a refinement as well as to exclude any possible temperature induced structural changes we performed single crystal XRD down to 100 K. Atomic coordinates, equivalent and anisotropic displacement as well as occupational parameters for the crystal structures of Sc$_5$Rh$_6$Sn$_{18}$ at 100 K, 150 K, 200 K, 250 K and RT are given in Tables\ \ref{tbl:crys3}, \ref{tbl:crys4}. In the whole studied temperature range the same structural model, which is now characterized by reasonable thermal anisotropic displacements and stoichiometric 5:6:18 composition, was refined. The performed refinements confirm Sc$_5$Rh$_6$Sn$_{18}$ to crystallize with Sc$_5$Ir$_6$Sn$_{18}$ structure type\cite{Levy2019}. In agreement with this finding, all reflections in the powder XRD patterns of Sc$_5$Rh$_6$Sn$_{18}$ were indexed with the same structural model. The unit cell parameters smoothly increase with increasing temperature (Fig.\ \ref{fig:latticeparam}) and the relative thermal expansion $\eta\, \approx\, 0.5$\, \% in the 100-300\, K range is typical for intermetallic compounds\cite{Kittel86}.

Interatomic distances in the crystal structures of Sc$_5$Rh$_6$Sn$_{18}$ refined with both Tb$_5$Rh$_6$Sn$_{18}$ and Sc$_5$Ir$_6$Sn$_{18}$ types models are given in Table\ \ref{tbl:dist}. They agree mainly well with the sums of atomic radii of the elements ($r_\mathrm{Sc}$ = 1.61 \AA, $r_\mathrm{Rh}$ = 1.34 \AA, $r_\mathrm{Sn}$ = 1.41 \AA\cite{Emsley1998}). In both models Sc2-Rh and Sc2-Sn contacts are almost equal with the corresponding sums, while the Sn-Sn distances are by $\sim$1-2 \% longer. The shrinking of Rh-Sn bonds is of $\sim$4.8 \% or $\sim$5.4 \% in the Tb$_5$Rh$_6$Sn$_{18}$ and Sc$_5$Ir$_6$Sn$_{18}$ types models, respectively. Of special interest are Sc1-Sn and Sn1-Sn contacts: they exceed by $\sim$15 \% the corresponding sums of atomic radii in the Tb$_5$Rh$_6$Sn$_{18}$ type model (Tables\ \ref{tbl:dist}). These long distances could classify Sc$_5$Rh$_6$Sn$_{18}$ as a cage-compound\cite{Kase2012,Bhattacha2018}.  However, applying the Sc$_5$Ir$_6$Sn$_{18}$ model Sc1-Sn and Sn1-Sn distances in this compound would exceed the corresponding sums by only $\sim$6 \% and $\sim$10 \%, respectively. On the other hand, the Sn1-Sc2 contact is shorter by $\sim$2~\%, and thus the Sn1 atom is involved in covalent bonding (see discussion below). Therefore, Sc$_5$Rh$_6$Sn$_{18}$ cannot be considered as a cage-compound.

Since the only difference between Tb$_5$Rh$_6$Sn$_{18}$ and Sc$_5$Ir$_6$Sn$_{18}$ types of arrangements is the split of some crystallographic sites (Fig.\ \ref{fig:Baernighausen}), we used the Tb$_5$Rh$_6$Sn$_{18}$ one for the further structural description. The close relationship of the Tb$_5$Rh$_6$Sn$_{18}$ type arrangement with the primitive cubic Remeika Yb$_3$Rh$_4$Sn$_{13}$ prototype (it is already reflected in the relations between the unit cell parameters: $a_\mathrm{tetr}\, \approx a_\mathrm{cub}\sqrt{2}$; $c_\mathrm{tetr}\, \approx 3a_\mathrm{cub}$) has been widely discussed in the literature\cite{Remeika1980,Miraglia1987,gum2018Rem}. Both structures are characterized by corner sharing [RnSn$_6$] trigonal prismatic arrays and reveal distorted [$M$Sn$_{12}$] cuboctahedra for $M$ = Yb, Sc1 atoms (Fig.\ \ref{fig:str}a-d). 
The common blocks in both types are cuboctahedra (grey) condensed with 4 trigonal prisms (tan), which share their Sn-vertices (Fig.\ \ref{fig:str}a and c). However, whereas in Yb$_3$Rh$_4$Sn$_{13}$ the free space in-between the blocks is filled with ideal [Sn1Sn$_{12}$] icosahedra (red in Fig.\ \ref{fig:str}a) in Sc$_5$Rh$_6$Sn$_{18}$ these are now 16 vertices distorted Frank-Kasper\cite{Alvarez2005} [Sn1Sc2$_2$Sn$_{14}$] polyhedra (Fig. \ref{fig:Anis}c, Sn-Sn bonds are red and Sc-Sn - dark yellow in Fig.\ \ref{fig:str}c). Another crucial difference between these two types is that condensed cuboctahedra form in Yb$_3$Rh$_4$Sn$_{13}$ infinite columns along $b$-direction (grey and transparent in Fig.\ \ref{fig:str}b), while in the structure of Sc$_5$Rh$_6$Sn$_{18}$ they are separated from each other (grey and transparent in Fig.\ \ref{fig:str}d). One could also present the Remeika phase as alternating layers of blocks of cuboctahedra (transparent and grey) condensed with 4 differently oriented trigonal prisms (tan in the first layer and blue and tan in the second one) (Fig.\ \ref{fig:str}b). In contrary, three layers can be identified for Sc$_5$Rh$_6$Sn$_{18}$ (Tb$_5$Rh$_6$Sn$_{18}$-type) structural arrangement (Fig.\ \ref{fig:str}d). The first one (tan in Fig.\ \ref{fig:str}d) is identical with that of Yb$_3$Rh$_4$Sn$_{13}$, the second layer consists of distorted [RnSn$_6$] trigonal prisms (green in Fig.\ \ref{fig:str}d) which are separated from each other and condensed with cuboctahedra from the 1$^\mathrm{st}$ layer. Finally the third layer is again from Yb$_3$Rh$_4$Sn$_{13}$ type [contains trigonal prisms condensed with cuboctahedra (blue and grey, respectively in Fig.\ \ref{fig:str}d)]. However, it is shifted by 1/2 of the translation in a-direction compared to the 1st layer.

\subsubsection{Transmission electron microscopy}

\begin{figure}
	\includegraphics[width=\linewidth]{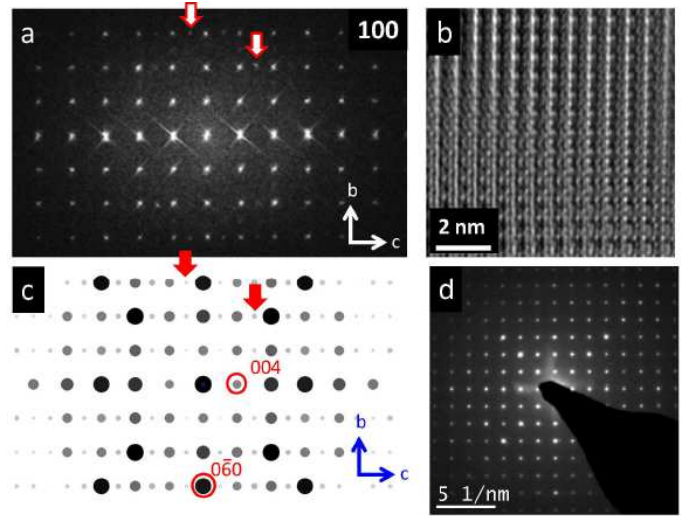}
	\caption{TEM of the [100] zone in Sc$_5$Rh$_6$Sn$_{18}$. (a) Fourier transform of the high-resolution image. Spots marked with arrows indicate (0\,6\,$\bar{2}$) and (0\,4\,6) reflections which are lacking in the [001] zone. (b) Corresponding Fourier filtered high-resolution image. (c) Simulated diffraction pattern of the [100] zone. (d) Electron diffraction pattern of the [100] zone. Intensities appear equalized due to dynamic scattering compared to the kinematic simulation in (c).}
	\label{fig:TEM100}
\end{figure}

Transmission electron microscopy (TEM) was performed on a single crystal of Sc$_5$Rh$_6$Sn$_{18}$ to verify the structure model derived from X-ray diffraction measurements and to check for the presence of the super structure reflections. A fast Fourier transform (FFT) of the [1\,0\,0] zone of the high-resolution image shows the expected, additional peaks e.g. (0\,6\,$\bar2$) or (0\,4\,6) which corroborate the tetragonal cell with $c$\,$\approx$\,$2a$ (Fig.\,\ref{fig:TEM100}). The electron diffraction pattern reveals rather equalized intensities of the reflections due to dynamic scattering (Fig.\,\ref{fig:TEM100}d). The (0\,6\,$\bar2$) or (0\,4\,6) reflections are also present in the diffraction pattern, but faintly. 
In accordance with the structure model from X-ray diffraction no superstructure reflections or streaking effects as reported for other MR$_6$Sn$_{18}$ (M=Gd, Tb, Dy, R=Rh, Os) \cite{Miraglia1987,Hodeau1982} were observed in the zones [0\,0\,1] and [1\,1\,0]. The electron diffraction patterns display good agreement with simulation (Fig.\,\ref{fig:TEM001}, \ref{fig:TEM110}).

The situation, where XRD refinement of large complex crystal structures results in numerous split positions and TEM reveals no indication of any superstructure could indicate local deviations from the translational symmetry and thus, appearance of some domains without inversion center. Such a local disorder with domains breaking the average symmetry is recently reported for Ba$_{7.81}$Ge$_{40.67}$Au$_{5.33}$ clathrate \cite{Lory2017}, boron carbide \cite{Rasim2018}, $\beta$-Mg$_2$Al$_3$ \cite{Samson1965,Feuerbacher2007} and Ruthenium Zinc Antimonides \cite{Xiong2010}. These findings were confirmed in the mentioned compounds by combined ab initio DFT-calculations and high resolution TEM. A local disorder assuming the presence of some non-centrosymmetric domains in the crystal structure of Sc$_5$Rh$_6$Sn$_{18}$ could support unconventional superconductivity mechanisms. On the other hand the rather complicated structure of Sc$_5$Rh$_6$Sn$_{18}$ could be supposed to smooth such local effects of missing inversion symmetry \cite{Bauer2012}.

\subsubsection{Electronic structure and chemical bonding situation}

To shed light on the possible changes in the chemical bonding situation caused by the shortening of some contacts due to many split of Wyckoff positions, electronic structure calculations were performed using both the Tb$_5$Rh$_6$Sn$_{18}$-type and a fictitious model, which was created based on the Sc$_5$Ir$_6$Sn$_{18}$ type. To create it the initial symmetry was lowered from SG $I4_1/acd$ to $I4_1cd$ ($i.e.$ the number of crystallographic sites was doubled) (Fig.\ \ref{fig:Baernighausen}) and, then the Wyckoff positions with the shortest contacts (found in Table\ \ref{tbl:dist}), were selected to obtain the composition Sc$_5$Rh$_6$Sn$_{18}$ with no splits. However, electronic structure and chemical bonding analysis revealed negligible differences between the models. Consequently, the features of the ideal Tb$_5$Rh$_6$Sn$_{18}$-type model will be presented and discussed here.

\begin{figure}
	\includegraphics[width=\linewidth]{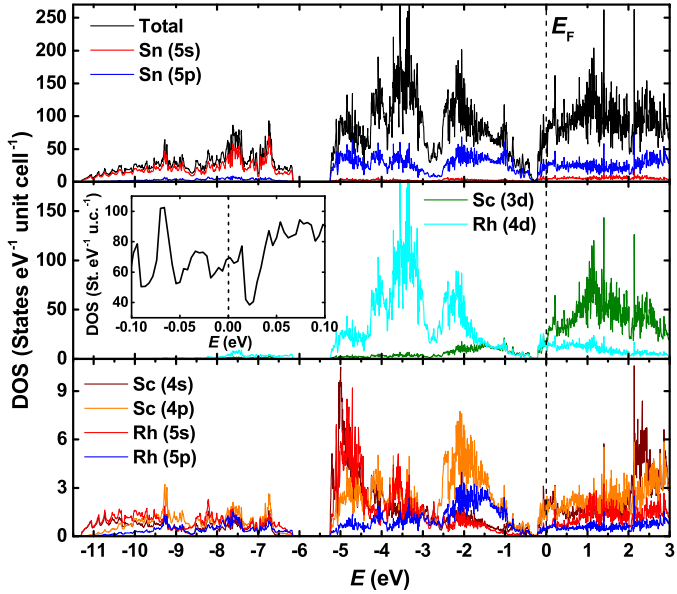}
	\caption{Electronic density of states (DOS) for Sc$_5$Rh$_6$Sn$_{18}$ assuming ideal  Tb$_5$Rh$_6$Sn$_{18}$-type model. Inset: Total DOS in the Energy range from -0.1--0.1 eV.}
	\label{fig:origDOS}
\end{figure}

The electronic density of states (DOS) for Sc$_5$Rh$_6$Sn$_{18}$ assuming the ideal Tb$_5$Rh$_6$Sn$_{18}$-type model is shown in Fig. \ref{fig:origDOS}. The value of the DOS at the Fermi energy (set to 0 eV) is 16.1 states eV$^{-1}$ prim.cell$^{-1}$. This indicates the theoretically calculated Sommerfeld coefficient of the electronic specific heat $\gamma_\mathrm{theor}$ = 38 mJ mol$^{-1}$ K$^{-2}$ to be by a factor of $\sim$2 smaller than the experimentally observed one \cite{Feig} and, thus an electronic instability of the Tb$_5$Rh$_6$Sn$_{18}$-type model. The states between -11.5 and -6 eV are heavily dominated by the 5$s$ contributions of the Sn atoms. The upper manifold of the valence region, [-5.25, 0] eV, consists mainly of Sn5$p$ and Rh$4d$ states. The $s$ and $p$ contributions from the Rh and Sc atoms are much smaller ($cf.$ bottom panel of Fig.\ \ref{fig:origDOS}). The lower part of the conduction band has largely Sc$3d$ and Sn$5p$ contributions. These observations are of general nature, and in order to have a deeper understanding of the bonding situation in Sc$_5$Rh$_6$Sn$_{18}$ position space chemical bonding analysis was carried out.

\begin{figure}
	\includegraphics[width=\linewidth]{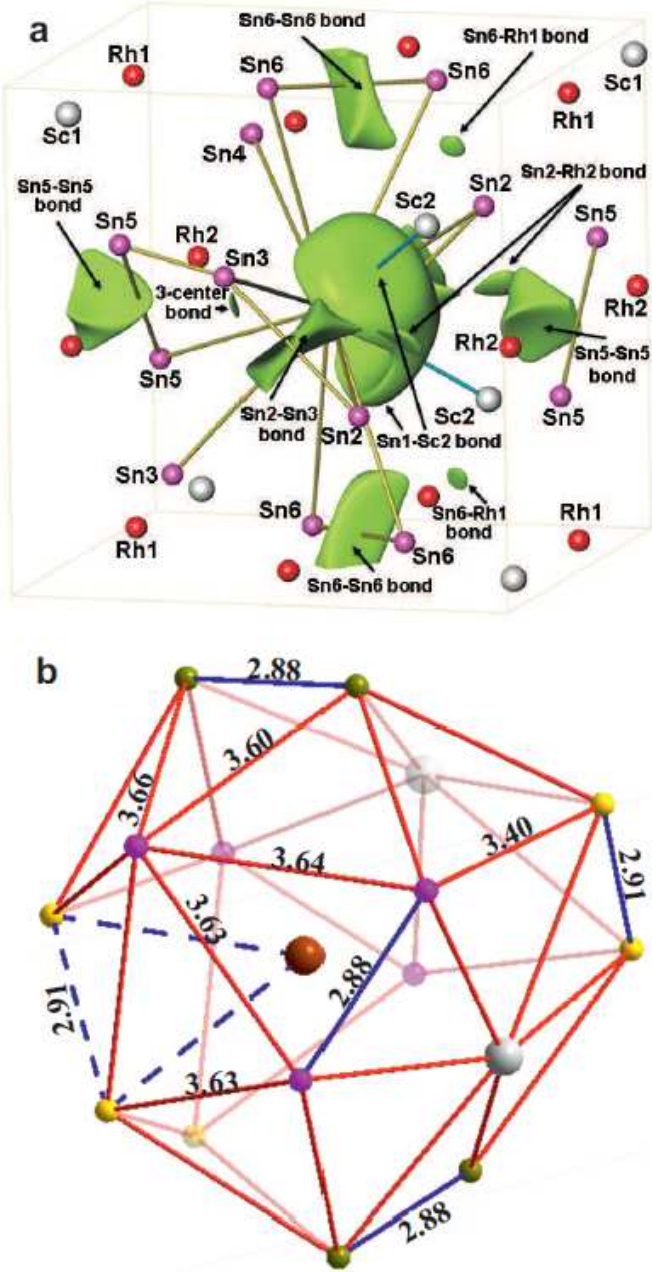}
	\caption{a: The ELI distribution around Sn1 at the center of the distorted Frank-Kasper polyhedron. The isosurface value is 1.0692. The atom types and bonds are identified. The thin orange lines show the boundaries of the box in which the ELI was computed. The location of the Sn1 atom is covered by the isosurface due to the Sn1 - Sc2 bond. The 3-center bond is formed by Sn1 and two Sn4 atoms. All Sn-Sn two-center bond types are shown. b: The structural arrangement in the same region as the ELI distribution illustrated by the [Sn1Sc2$_2$Sn$_{14}$] polyhedron. The distances between the respective Sn atoms are rounded to two digits and given in \AA. The bonds discussed in the text are colored in blue. The Sn1 and Sc2 atoms are illustrated as large brown or gray spheres, respectively. The smaller spheres represent Sn2-Sn4 (pink), Sn5 (yellow) and Sn6 (green).}
	\label{fig:ELI}
\end{figure}

The topological analysis of the ED yields the atomic or QTAIM basins \cite{Quantumtheory}. Integration of the ED inside these basins gives the electron population of each atom, from which effective charges can be calculated. In Sc$_5$Rh$_6$Sn$_{18}$ as expected Sc (Rh) atoms are positively (negatively) charged, 1.3 (-1.2) (see Tables \ref{tbl:qtaim} and \ref{tbl:effchargecompar}). The results for the Sn atoms are mixed: the effective charge of Sn1 is -0.5, while all the other Sn atoms are positively charged. Note that the values for Sn2 (at \textit{16f}), Sn5 and Sn6 (both at the general position $32g$) are less than 0.1. The negative charge on Sn1 can be attributed to the absence of Rh atoms (most electronegative element in this compound) in its first coordination shell.

\begin{table}
	\caption{Effective charges obtained through QTAIM analysis}
	\centering
	\label{tbl:qtaim}
	\begin{tabular}{c@{\hspace{1cm}}c}
		\hline
		Wyckoff position 						& Effective Charge 					\\
		\hline
		Sc1									    & +1.32						       \\
		Sc2   								    & +1.30						       \\
		Rh1    									& -1.18								\\
		Rh2										& -1.17           					\\
		Sn1  									& -0.53 							\\
		Sn2 									& -0.05  					        \\
		Sn3 									& +0.14					            \\
		Sn4               						& +0.18							    \\
		Sn5                 					& +0.06						        \\
		Sn6            							& +0.06          					\\
\hline
	\end{tabular}
\end{table} 

The main features of the ELI-D topological analysis in Sc$_5$Rh$_6$Sn$_{18}$ can be summarized as one type of three-center Sn-only bonds and various two-center bonds formed by Sn-Sn, Sn-Sc, Sn-Rh and Rh-Sc pairs. In all cases the participating atoms are near neighbors. The two-center Sn-Sn bonds involve Sn5-Sn5 (1.98 e$^-$), Sn6-Sn6 (2.16 e$^-$) and Sn2-Sn3 (2.14 e$^-$) atoms having distances shorter than 2.91~\AA ~(see Table\ \ref{tbl:dist}). The three-center bond is formed by Sn1-Sn5-Sn5 and the Sn1-Sn5 distance, 3.244~\AA, is the fourth shortest Sn-Sn contact in this compound (Fig. \ref{fig:ELI}b). Sn1 contributes 70~\% of the total bond electrons, $\sim$1.95. Only one type of Sn-Sc contact is shorter than 3~\AA, between Sn1 and Sc2. Thus, these atoms form the only Sn-Sc bond with a high bond polarity, contribution of Sc2 is about 7~\% of the total bond electrons which is 1.9 (Fig. \ref{fig:ELI}a). The closest Sn neighbors of Rh1 are Sn4, Sn5 and Sn6. Electron populations of the corresponding bonds are between 1.72, 1.91 and 1.92, respectively. Rh1 atoms contribute $\sim$25~\% of them. The near neighbours of Rh2 atoms include all Sn atoms but Sn1. Bond electron populations vary between 1.7 and 2.0 with Rh2 accounting for 22 -- 26~\%. The Sc2-Rh1 and Sc2-Rh2 bonds are also polar with Sc2 contributions at the level of about 17~\%. However, the basins of these bonds contain fewer electrons, $\sim$0.3. The distances in the first cuboctahedral coordination shell of Sc1 are all longer than 3.34~\AA ~(see Table\ \ref{tbl:dist}), hence there are no two-center bonds involving Sc1 atoms as a major participant; Sc1 contributes to various two-center bonds only at few percent level. Therefore, Sc1 atoms' participation in atomic interactions is largely of ionic nature.

\begin{figure}
	\includegraphics[width=\linewidth]{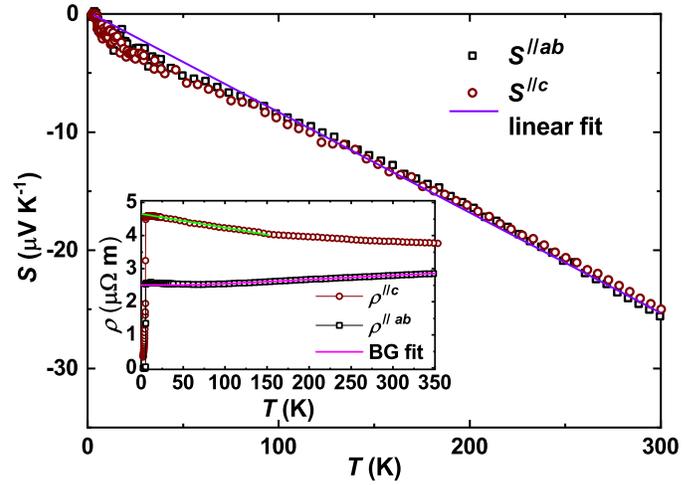}
	\caption{Temperature dependence of the thermopower $S(T)$ for Sc$_5$Rh$_6$Sn$_{18}$. The violet line represents a linear fit of the data to $S = \alpha T$ in the temperature range 200\,K\,-\,350\,K.}
	\label{fig:s}
\end{figure}

\begin{figure}[h]
	\includegraphics[width=\linewidth]{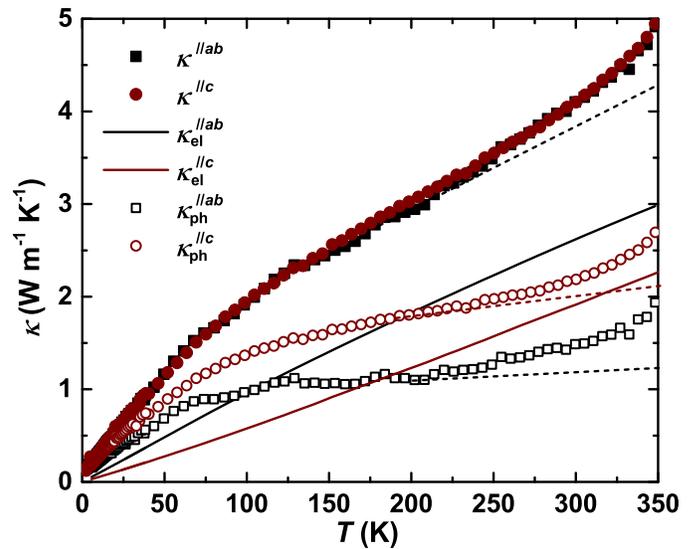}
	\caption{(Color online) Temperature dependence of thermal conductivity $\kappa$($T$) of Sc$_5$Rh$_6$Sn$_{18}$ with its electronic ($\kappa_{\mathrm{el}}$) and phonon ($\kappa_{\mathrm{ph}}$) contributions measured in different directions. In the high-temperature range above 200 K, ($\kappa_{\mathrm{exp}}$) and ($\kappa_{\mathrm{el}}$) for Sc$_5$Rh$_6$Sn$_{18}$ are biased due to radiation heat losses which follow roughly a $\propto T^3$ law. The expected real conductivities are shown by dotted lines.}
	\label{fig:kappa}
\end{figure}

\subsection{Electrical and Thermal transport properties}\label{transport}

The electrical resistivity of Sc$_5$Rh$_6$Sn$_{18}$ reveals an anisotropic bad metallic behavior above the superconducting transition (inset to Fig. \ref{fig:s}). It increases with increasing temperature in the direction parallel to \textit{ab} and almost linearly decreases for $\rho \parallel c$. Such a temperature dependence of $\rho$(\textit{T}) could be ascribed to the strong structural disorder and low charge carrier concentration observed in Sc$_5$Rh$_6$Sn$_{18}$ (for more details see discussion in \cite{Feig})

The temperature dependence of the thermopower $S(T)$ of Sc$_5$Rh$_6$Sn$_{18}$ measured along $ab$ ($S \parallel ab$)- and $c$ ($S \parallel c$)-directions is shown in Fig.\,\ref{fig:s}. No anisotropy is observed. $S$($T$) is negative in the whole measured temperature range (1.8\,K--400\,K) and indicates the dominance of electron-like charge carriers in agreement with reported Hall-effect studies \cite{scte2013}. The thermopower in the temperature range 200\,K $\leq\ T \leq$ 350\,K is described by $S = \alpha T$ with a small slope $\alpha$\,=\,$-0.08$\,$\mu$V K$^{-2}$. At lower temperatures the experimental curves deviate slightly from the linear behavior due to the presence of additional scattering mechanisms. Such behavior of $S$($T$) resembles a typical metal-like behavior. The dimensionless parameter $q = N_Ae\alpha/\gamma$ (Faraday constant $N_A \,e \approx 96485$ C\,mol$^{-1}$), which characterizes the thermoelectric material in terms of an effective charge carrier
concentration per f.u.\ (or the Fermi volume $V_\mathrm{F}$ of the
charge carriers) \cite{Zlatic2007}, is estimated to be 0.11 for
Sc$_5$Rh$_6$Sn$_{18}$. For Fermi liquids the value of $q$ is expected to be approximately 1.

The temperature dependence of the thermal conductivity $\kappa$($T$) for Sc$_5$Rh$_6$Sn$_{18}$ with the heat flow parallel to the $ab$ ($\kappa \parallel ab$)- and  $c$ ($\kappa \parallel c$)-directions is plotted in Fig.\ \ref{fig:kappa}. With decreasing temperature $\kappa(T)$ decreases in the whole measured temperature range showing a small kink at $T_\mathrm{kink}$ = 75\,K. No anisotropic behavior of the thermal conductivity within the standard deviation was observed. The total $\kappa$($T$) is small over the whole measured temperature range, which is obviously due to the strong structural disorder in Sc$_5$Rh$_6$Sn$_{18}$.

For metallic compounds $\kappa(T)$ can be decomposed into
$\kappa_\mathrm{el}$ and $\kappa_\mathrm{ph}$. The electronic contribution $\kappa_\mathrm{el}$ may be estimated from the experimental resistivity data by applying the Wiedemann-Franz law $\kappa_{\mathrm{el}}(T) = L_0T/\rho(T)$, where $L_0$ is the Lorenz number $2.44 \times 10^{-8}$ W\,$\Omega$\,K$^{-2}$. The phonon contribution $\kappa_\mathrm{ph}$ for Sc$_5$Rh$_6$Sn$_{18}$ was estimated by subtracting $\kappa_\mathrm{el}$ from the measured value of $\kappa$. Due to the anisotropy in the electrical resistivity $\kappa_\mathrm{el}^{\parallel ab}$ is dominating the thermal conductivity for $T > 100$\, K, while for the $c$-direction the main contribution is from $\kappa_\mathrm{ph}^{\parallel c}$.

\section{Conclusions}

Crystal structure refinements performed for Sc$_5$Rh$_6$Sn$_{18}$ in the temperature range of 100-300\, K from both high-resolution synchrotron powder and single crystal X-ray diffraction methods revealed it to crystallize with the centrosymmetric Sc$_5$Ir$_6$Sn$_{18}$ structure type (a split variant of the Tb$_5$Rh$_6$Sn$_{18}$ prototype). Such a structural model assumes shortening of Sc-Sn contacts and does not allow to classify Sc$_5$Rh$_6$Sn$_{18}$ as a cage-compound. Further TEM characterization confirmed the absence of any satellites (i.e. possible modulation) or superstructure reflections (i.e. changes in unit cell dimensions). Splits of numerous Sn-positions together with the absence of any signatures of superstructure can however indicate a local disorder and thus, the possible absence of an inversion center in some domains. They can be indicative of packing incompatibilities of the large polyhedra. Both Sc$_5$Ir$_6$Sn$_{18}$ and Tb$_5$Rh$_6$Sn$_{18}$ types show close relationships with the primitive cubic Yb$_3$Rh$_4$Sn$_{13}$ Remeika phase revealing a similar array of corner sharing [RhSn$_6$] trigonal prisms condensed with distorted [$M$Sn$_{12}$] ($M$ = Yb or Sc) cuboctahedra.   

The performed chemical bonding situation analysis revealed highly polar character for Sc2-Sn1, Sn-Rh and Sc2-Rh bonds as well as two- and three-center bonds involving Sn-atoms. Sc1 atoms' participation in atomic interactions is mainly of ionic nature. No differences in  bonding situations between the idealized Tb$_5$Rh$_6$Sn$_{18}$-like model and a fictitious structure, which was created based on the Sc$_5$Ir$_6$Sn$_{18}$ type and included only the shortest contacts, was observed. 

Thus, both crystal structure refinements as well as analysis of chemical bonding situation confirm the average crystal structure of Sc$_5$Rh$_6$Sn$_{18}$ to be centrosymmetric, which would be rather incompatible with unconventional superconductivity (SC) with a non unitary triplet electron-phonon pairing with point nodes in the gap as recently proposed \cite{Bhattacha2018}. On the other hand the local disorder could support such a scenario. This prompted us to perform an additional study of SC properties of Sc$_5$Rh$_6$Sn$_{18}$, which revealed it to be a conventional SC with stronger electron-phonon coupling and isotropic $s$-wave gap \cite{Feig}.

The anisotropic behavior of temperature dependence of electrical resistivity in Sc$_5$Rh$_6$Sn$_{18}$ is in details discussed in \cite{Feig}. The thermopower $S(T)$ of the stannide reveals no anisotropy and is small and negative in the whole measured temperature range, which indicates dominance of electron-like charge carriers in this metallic system.

The thermal conductivity $\kappa$($T$) for  Sc$_5$Rh$_6$Sn$_{18}$ is isotropic as well and smaller than for a metal. This can be explained with the intrinsic structural disorder in the studied compound. In accordance with the anisotropy in the electrical resistivity $\kappa_\mathrm{el}^{\parallel ab}$ is dominating thermal conductivity for $T > 100$\, K, while for the $c$-direction the main contribution is from $\kappa_\mathrm{ph}^{\parallel c}$.  

\section*{Acknowledgments}

This work is performed within the DFG (Deutsche Forschungsgemeinschaft) grant 325295543. The authors are grateful to J.\ Grin for his interest and steady support. The authors thank U.\ Burkhardt, P.\ Scheppan and S.\ Kostmann for the metallographical analysis, J.\ Chang for his contribution in the initial phase of this project. We acknowledge D.\ Sokolov for providing a high quality Laue pattern of the crystal. H.\ Borrmann contributed single crystal XRD measurements at the initial stage of this research. The authors are indebted C.\ Hennig for his support during the adjustment of BM20 beamline at ESRF.

\clearpage

\appendix
\onecolumn
\section*{Supporting Information}

\renewcommand{\thefigure}{S\arabic{figure}}
\setcounter{figure}{0}   

\renewcommand{\thetable}{S\arabic{table}}
\setcounter{table}{0}   

\subsection{Tables}
\ \\
\ \\
\ \\
\ \\
\ \\
\ \\
\ \\
\ \\
\ \\
\ \\
\begin{table*}[h]
	\caption{Crystallographic data for Sc$_5$Rh$_6$Sn$_{18}$ at room temperature [SG \textit{I}4$_\mathrm{1}/acd$, \textit{Z} = 8, \textit{a} = 13.5529(2) \AA, \textit{c} = 27.0976(7) \AA] measured on crystals after syntheses from Sn-flux (as-cast), additionally stress annealed (at 1070 K for 2 h) and on the powdered crystal.}
	\label{tbl:struct_rt}
	\begin{tabularx}{\textwidth}{XXXX}
		\hline
		Material 	& as-cast crystal	& stress annealed crystal 		& powder			\\
		Diffractometer	&BM01 	& Rigaku AFC7 		& BM20 			\\
		Radiation, $\lambda$ (\AA)& 0.73331 	& Mo$K_\mathrm{\alpha}$ 0.71073	 & 0.45920 	\\ 
		Calculated density $\rho$ (g cm$^{-3}$)	& \multicolumn{2}{c}{7.94(2)}		& 7.65(1)		\\ 
		Crystal size ($\mu$m)	& 70 $\times$ 60 $\times$ 80	& 50 $\times$ 30 $\times$ 60 & polycrystalline	\\ 
		Maximum 2$\theta$($^{\circ}$) 	& 64.33	& 67.25 & 28.17    	\\ 
		\multirow{3}{\columnwidth}{Ranges in $h, k, l$}	& -17 $\leq$ \textit{h} $\leq$ 17	& -20 $\leq$ \textit{h} $\leq$ 10 & 0 $\leq$ \textit{h} $\leq$ 14	\\ 
			& -18 $\leq$ \textit{k} $\leq$ 18	& -17 $\leq$ \textit{k} $\leq$ 21 & 0 $\leq$ \textit{k} $\leq$ 20	\\ 
			& -39 $\leq$ \textit{l} $\leq$ 39	& -41 $\leq$ \textit{l} $\leq$ 27 & 0 $\leq$ \textit{l} $\leq$ 40	\\ 
		Absorption correction 	& \multicolumn{2}{c}{multi-scan} & -	\\
		\textit{T}(max)/\textit{T}(min) 	& 1.00/0.81	& 1.00/0.66 & -	\\
		Absorption coeff. (mm$^{-1}$)&30.12&22.71&15.73\\
		\textit{N}(hkl) measured&16794&19662&-\\
		\textit{N}(hkl) unique&1862&1863&-\\
		$R_\mathrm{int}$&0.015&0.040&-\\
		\textit{N}(hkl) observed&15204&18109&780\\
		Observation criterion&\multicolumn{2}{c}{\textit{F}(\textit{hkl}) $\geq$ 4$\sigma (F)$}&-\\
		Refined parameters&\multicolumn{2}{c}{70}&31\\
		$R_\mathrm{F}$, $R_\mathrm{I}$; $R_\mathrm W$, $R_\mathrm P$&0.020; 0.021&0.034; 0.036&0.056; 0.118\\
		Residual peaks (e \AA$^{-3}$)&-0.42/0.69&-0.83/1.00&-1.46/1.88\\
		\hline
	\end{tabularx}
\end{table*}

\clearpage

\begin{table*}
	\caption{Atomic coordinates, occupational parameters (\textit{G}), equivalent (Beq) and isotropic thermal displacement parameters for Sc$_\mathrm{5}$Rh$_\mathrm{6}$Sn$_\mathrm{16}$ at room temperature refined for crystals after syntheses from Sn-flux (as-cast) (1), additionally stress annealed (at 1070 K for 2 h) (2) and for a powdered crystal (3).}
	\label{tbl:powder-nosplit}
	\begin{tabularx}{\textwidth}{XXXXXXXX}
		Atom & Site & Material & \textit{G} & \textit{x}	& \textit{y} & \textit{z}&$B_\mathrm{eq}$/$B_\mathrm{iso}$ \\
		\hline
		\multirow{3}{\columnwidth}{Sc1}&\multirow{3}{\columnwidth}{8b}&1&0.92(1)&\multirow{3}{\columnwidth}{0}&\multirow{3}{\columnwidth}{1/4}&\multirow{3}{\columnwidth}{1/8}&3.06(3) \\ &&2&0.93(1)&&&&3.55(5)\\&&3&0.91(1)&&&&1.54(1) \\
		\hline
		\multirow{3}{\columnwidth}{Sc2}&\multirow{3}{\columnwidth}{32g}&1&\multirow{3}{\columnwidth}{-}&0.13321(3)&0.11563(3)&0.30808(1)&0.770(7)\\&&2&&0.13310(5)&0.11577(5)&0.30814(2)&0.83(2)\\&&3&&0.1323(9)&0.1196(9)&0.3166(5)&1.83(1) \\
		\hline
		\multirow{3}{\columnwidth}{Rh1}&\multirow{3}{\columnwidth}{16d}&1&\multirow{3}{\columnwidth}{-}&\multirow{3}{\columnwidth}{0}&\multirow{3}{\columnwidth}{1/4}&0.25363(1)&0.607(4)\\&&2&&&&0.25375(1)&0.652(6)\\&&3&&&&0.2555(5)&1.77(1)\\
		\hline
		\multirow{3}{\columnwidth}{Rh2}&\multirow{3}{\columnwidth}{32g}&1&\multirow{3}{\columnwidth}{-}&0.25761(1)&0.25003(1)&0.12515(1)&0.589(3)\\&&2&&0.25782(2)&0.24999(2)&0.12515(2)&0.648(5)\\&&3&&0.2574(4)&0.2507(3)&0.1223(2)&1.76(1) \\
		\hline
		\multirow{3}{\columnwidth}{Sn1}&\multirow{3}{\columnwidth}{16e}&1&\multirow{2}{\columnwidth}{-}&0.28054(2)&\multirow{3}{\columnwidth}{0}&\multirow{3}{\columnwidth}{1/4}&1.805(5)\\&&2&&0.28055(3)&&&1.902(9)\\&&3&0.90(1)&0.2728(3)&&&2.70(1)\\
		\hline
		\multirow{3}{\columnwidth}{Sn2}&\multirow{3}{\columnwidth}{16f}&1&\multirow{3}{\columnwidth}{-}&0.32507(1)&\multirow{3}{\columnwidth}{x+1/4}&\multirow{3}{\columnwidth}{1/8}&1.932(5)\\&&2&&0.32525(2)&&&1.828(8)\\&&3&&0.3237(4)&&&1.64(1)\\
		\hline
		\multirow{3}{\columnwidth}{Sn3}&\multirow{3}{\columnwidth}{16f}&1&\multirow{3}{\columnwidth}{-}&0.17545(1)&\multirow{3}{\columnwidth}{x+1/4}&\multirow{3}{\columnwidth}{1/8}&2.090(5)\\&&2&&0.17533(2)&&&1.957(8)\\&&3&&0.1756(4)&&&1.55(1)\\
		\hline
		\multirow{3}{\columnwidth}{Sn4}&\multirow{3}{\columnwidth}{32g}&1&\multirow{2}{\columnwidth}{-}&0.08756(1)&0.16251(1)&0.41897(1)&0.950(3)\\&&2&&0.08759(2)&0.16247(2)&0.41896(1)&1.034(5)\\&&3&0.94(1)&0.0874(4)&0.1636(4)&0.4183(2)&2.52(1)\\
		\hline
		\multirow{3}{\columnwidth}{Sn5}&\multirow{3}{\columnwidth}{32g}&1&\multirow{2}{\columnwidth}{-}&0.17450(1)&0.25582(2)&0.03775(1)&1.599(4)\\&&2&&0.17438(2)&0.25574(2)&0.03784(1)&1.565(7)\\&&3&0.97(1)&0.1706(7)&0.2541(3)&0.0379(4)&2.56(1)\\
		\hline
		\multirow{3}{\columnwidth}{Sn6}&\multirow{3}{\columnwidth}{32g}&1&\multirow{2}{\columnwidth}{-}&0.00597(2)&0.07421(1)&0.03749(1)&2.071(4)\\&&2&&0.00560(3)&0.07440(2)&0.03761(2)&1.972(8)\\&&3&0.91(1)&0.0030(3)&0.0780(7)&0.0374(4)&2.56(1)\\
		\hline
	\end{tabularx}
\end{table*} 

\clearpage

\begin{table*}
	\caption{Anisotropic thermal displacement parameters for Sc$_\mathrm{5}$Rh$_\mathrm{6}$Sn$_\mathrm{18}$ refined with the Tb$_\mathrm{5}$Rh$_\mathrm{6}$Sn$_\mathrm{18}$ model for crystals after syntheses from Sn-flux (as-cast) (1) and additionally stress annealed (at 1070 K for 2 h) (2).}
	\label{tbl:bvalues}
	\begin{tabularx}{\textwidth}{XXXXXXXX}
		Atom&Material&$B_\mathrm{11}$&$B_\mathrm{22}$&$B_\mathrm{33}$&$B_\mathrm{12}$&$B_\mathrm{13}$&$B_\mathrm{23}$\\
		\hline
		\multirow{2}{\columnwidth}{Sc1}&1&2.87(4)&$B_\mathrm{11}$&3.62(6)&0.26(4)&0&0\\
		&2&3.29(6)&$B_\mathrm{11}$&4.06(1)&0.21(8)&0&0\\
		\hline
		\multirow{2}{\columnwidth}{Sc2}&1&0.70(2)&0.76(2)&0.85(1)&0.04(1)&0.02(1)&0.05(1)\\
		&2&0.74(3)&0.84(3)&0.92(2)&0.01(2)&0.04(2)&0.07(2)\\
		\hline
		\multirow{2}{\columnwidth}{Rh1}&1&0.514(8)&0.537(8)&0.769(7)&-0.039(5)&0&0\\
		&2&0.57(1)&0.57(1)&0.82(1)&-0.030(9)&0&0\\
		\hline
		\multirow{2}{\columnwidth}{Rh2}&1&0.648(6)&0.497(6)&0.622(5)&-0.004(4)&0.019(5)&0.045(4)\\
		&2&0.703(8)&0.573(8)&0.668(8)&-0.025(7)&0.01(2)&0.054(7)\\
		\hline
		\multirow{2}{\columnwidth}{Sn1}&1&0.908(8)&2.27(1)&2.228(8)&0&0&-1.414(7)\\
		&2&1.00(1)&2.40(2)&2.31(2)&0&0&-1.40(2)\\
		\hline
		\multirow{2}{\columnwidth}{Sn2}&1&0.787(5)&$B_\mathrm{11}$&4.23(1)&0.254(6)&-0.294(5)&-$B_\mathrm{13}$\\
		&2&0.822(8)&$B_\mathrm{11}$&3.84(2)&0.27(1)&-0.24(1)&-$B_\mathrm{13}$\\
		\hline
		\multirow{2}{\columnwidth}{Sn3}&1&0.751(5)&$B_\mathrm{11}$&4.77(2)&0.261(6)&-0.165(5)&-$B_\mathrm{13}$\\
		&2&0.820(8)&$B_\mathrm{11}$&4.23(2)&0.28(1)&-0.15(1)&-$B_\mathrm{13}$\\
		\hline
		\multirow{2}{\columnwidth}{Sn4}&1&0.910(6)&0.900(6)&1.037(5)&-0.380(4)&0.378(4)&-0.387(4)\\
		&2&1.006(9)&0.983(9)&1.113(8)&-0.389(6)&0.395(7)&-0.382(7)\\
		\hline
		\multirow{2}{\columnwidth}{Sn5}&1&0.853(6)&2.956(8)&0.983(5)&0.387(5)&-0.331(4)&-0.362(5)\\
		&2&0.89(1)&2.79(2)&1.01(2)&0.343(8)&-0.323(8)&-0.325(8)\\
		\hline
		\multirow{2}{\columnwidth}{Sn6}&1&4.54(1)&0.788(6)&0.881(6)&-0.411(6)&0.097(5)&0.261(4)\\
		&2&4.14(2)&0.87(1)&0.91(2)&-0.34(1)&0.097(9)&0.279(9)\\
		\hline
	\end{tabularx}
\end{table*} 

\clearpage

\begin{landscape}
	\begin{table}
		\caption{Crystallographic data for Sc$_5$Rh$_6$Sn$_{18}$ as obtained from synchrotron single crystal diffraction data in the range 100\,K to 300\,K.}
		\label{tbl:crys3}
		\begin{tabular}{l@{\hspace{0.1cm}}l@{\hspace{0.1cm}}l@{\hspace{0.1cm}}l@{\hspace{0.1cm}}l@{\hspace{0.1cm}}l}
			\hline	
				& 100\,K  & 150\,K  	&200\,K 	 &250\,K 	 &300\,K  \\
			\hline
				$R_F$/$R_w$ 	& 	2.1/2.2\,\%		&		2.1/2.2\,\%	&	2.3/2.4\,\%		&	2.71/2.8\,\%	&	2.2/2.3\,\%  \\
				Sc1 $x$;$y$; & -0.0011(8);0.2420(6);& -0.0010(10);0.2437(8)&-0.0020(11);0.2435(9);&-0.002(2);0.2443(15);&-0.002(2);0.244(2);\\
				$z$;$B_\mathrm{eq}$;\textit{G}&0.1192(2);1.9(2);0.25 &0.1199(2);2.2(4);0.25	&0.1201(2);2.4(5);0.25&0.1216(5);2.9(7);0.25&0.1212(5);3.0(7);0.25\\
				Sc2 $x$;$y$;&	0.13328(7);0.11535(7)	 &0.13327(7);0.11544(7);&0.13319(8);0.11555(8);	&0.1331(1);0.1156(1);&0.1331(1);0.1158(1);\\
				$z$;$B_\mathrm{eq}$;\textit{G}&0.30805(4);0.42(2);1 &0.30803(4);0.49(2);1&0.30810(4);0.57(2);1	&0.30814(5);0.68(3);1&0.30818(5);0.73(3);1 \\
				Rh1 $x$;$y$; &0;0.25; &	0; 0.25; &0;0.25;&	0;0.25;&0;0.25; \\
				$z$;$B_\mathrm{eq}$;\textit{G}&0.99644(2);0.35(1);1 &0.99644(2);0.40(1);1	&0.99642(2);0.48(1)&0.99643(3);0.52(2);1&0.99643(3);0.73(3);1 \\
				Rh2 $x$;$y$; &-0.00011(3);0.99253(3) &	-0.00009(3);0.99251(3);	&-0.00006(3);0.99246(3);&-0.00002(4);0.99246(4);&-0.00001(4);0.99248(4);\\
				$z$;$B_\mathrm{eq}$;\textit{G}&0.12484(1);0.328(7);1 &	0.12484(1);0.386(7);1&0.12485(2);0.458(8);1&0.12485(2);0.51(1);1&0.12485(2);0.54(1);1\\
				Sn1 $x$;$y$;&0.24254(7);0.03079(4);	 &0.24219(7);0.03073(4);&0.24184(7);0.03070(5);&0.24126(9);0.03071(6);&0.2414(1);0.03063(6);\\
				$z$;$B_\mathrm{eq}$;\textit{G}&0.49628(4);0.44(1);0.5 &0.49616(4);0.55(1);0.5&0.49597(4);0.68(2);0.5&0.49604(5);0.80(2);0.5&0.49568(5);0.88(2);0.5\\
				Sn2 $x$;$y$; & 0.17380(6);0.07379(6); &	0.17383(6);0.07384(6);	&0.17383(7);0.07396(7);&0.17386(9);0.07393(9);&0.17398(9);0.07401(9);\\
				$z$;$B_\mathrm{eq}$;\textit{G}& 0.11812(3);0.59(2);0.5 &0.11826(3);0.66(2);0.5&0.11839(3);0.73(2);0.5&0.11863(4);0.84(3);0.5&0.11851(4);0.86(3);0.5\\
				Sn3 $x$;$y$;&	0.17529(6);0.07374(6); &0.17522(6);0.07382(6)&0.17511(6);0.07389(6);&0.17502(8);0.07390(8);&0.17498(9);0.07394(9);\\
				$z$;$B_\mathrm{eq}$;\textit{G}&0.63223(3);0.60(2);0.5 &0.63211(3);0.66(2);0.5&	0.63203(3);0.73(2);0.5&0.63183(3);0.86(3);0.5&0.63193(4);0.87(3);0.5\\
				Sn4 $x$;$y$; & 0.08705(3);0.33706(3); &	0.08716(3);0.33716(3);	&0.08730(3);0.33731(3);&0.08741(4);0.33743(4);&0.08750(4);0.33755(4);\\
				$z$;$B_\mathrm{eq}$;\textit{G}&0.33127(1);0.496(7);1 &0.33122(1);0.597(7);1&0.33116(2);0.67(2);1&0.33112(2);0.81(1);1&0.33106(2);0.88(1);1\\
				Sn5a $x$;$y$;	& 0.17704(6);0.26565(5); &0.17686(6);0.26572(5);&0.17677(7);0.26580(5);&0.17666(8);0.26605(7);& 0.17662(10);0.26600(8);\\
				$z$;$B_\mathrm{eq}$;\textit{G}& 0.03661(3);0.52(2);0.518(2) &0.03667(3);0.58(2);0.518(2)&0.03670(3);0.67(2);0.518(3)&0.03681(4);0.72(2);0.518(3)&0.03672(5);0.79(3);0.518(3)\\
				Sn5b $x$;$y$; & 0.00520(6);0.07844(7); &	0.00537(6);0.07834(7);&0.00547(7);0.07818(8);&0.00592(8);0.07802(9);&0.00562(9);0.0779(1);\\
				$z$;$B_\mathrm{eq}$;\textit{G}& 0.21086(4);0.56(2);0.482(2) &0.21090(4);0.63(2);0.482(2)&0.21096(4);0.73(2);0.482(3)&0.21110(5);0.78(3);0.482(3)&0.21105(5);0.88(3);0.482(3)\\	
				Sn6a $x$;$y$;&0.02212(6);0.07198(7);	 &0.02176(6);0.07214(7);&0.02132(6);0.07230(7);&0.02096(8);0.07245(9);&0.02029(9);0.07255(10);\\
				$z$;$B_\mathrm{eq}$;\textit{G}&0.03744(3);0.65(2);0.482(2)  &0.03748(3);0.68(2);0.482(2)&0.03750(4);0.74(2);0.482(3)&0.03754(4);0.78(3);0.482(3)&0.03755(5);0.89(3);0.482(3) \\	
				Sn6b $x$;$y$; & 0.99402(6);0.07556(6); &	0.99387(6);0.07560(6);&	0.99368(6);0.07558(6);&0.99335(7);0.07550(8);& 0.99301(8);0.07557(9);\\
				$z$;$B_\mathrm{eq}$;\textit{G}& 0.03748(3); 0.59(2);0.518(2)&0.03746(3);0.66(2);0.518(2)&0.03670(3);0.67(2);0.518(3)&0.03745(4);0.82(2);0.518(3)&0.03743(4);0.89(3);0.518(3)\\
				\hline
		\end{tabular}
	\end{table} 
\end{landscape}

\begin{landscape}
	\begin{table}
		\caption{Anisotropic thermal displacement parameters for Sc$_5$Rh$_6$Sn$_{18}$ at different temperatures.}
		\label{tbl:crys4}
		\begin{tabular}{l@{\hspace{0.1cm}}l@{\hspace{0.1cm}}l@{\hspace{0.1cm}}l@{\hspace{0.1cm}}l@{\hspace{0.1cm}}r}
				\hline
				& 100\,K  & 150\,K  	&200\,K 	 &250\,K 	 &300\,K  \\
					\hline
				Sc1 B$_{11}$;B$_{22}$;B$_{33}$; &2.1(4);1.9(4);1.8(3) & 2.0(7);2.5(8);2.0(2); & 2.2(9);2.7(10);2.3(2); & 2.8(27);2.9(28);3.0(3); &2.6(13);3.4(16);3.0(3);\\
				B$_{12}$;B$_{13}$;B$_{23}$; &0.12(13);0.4(5);0.4(5) & 0.06(12);0.5(14);-0.4(14); &0.06(14);0.6(14);-0.3(14);  &0.1(2);0.3(27);0.0(25);  &0.0(2);0.5(18);-0.5(17);\\
				Sc2 B$_{11}$;B$_{22}$;B$_{33}$;&0.33(3);0.37(3)0.56(3) &0.41(3);0.45(3);0.62(3); & 0.50(4);0.55(4);0.67(4); & 0.65(5);0.70(5);0.69(4); &0.64(5);0.71(5);0.83(5);;\\
				B$_{12}$;B$_{13}$;B$_{23}$; &0.00(3);0.02(3);0.01(3) & -0.00(3);0.02(3);0.03(3); &-0.02(3);0.03(3);0.02(3);  & -0.00(4);0.00(4);0.04(4); &0.02(4);0.03(4);0.03(4);\\
				Rh1 B$_{11}$;B$_{22}$;B$_{33}$;&0.28(2);0.25(2);0.51(2) & 0.34(2);0.32(2);0.54(2); &0.43(2);0.41(2);0.59(2);  & 0.52(3);0.51(3);0.53(3); &0.49(3);0.47(3);0.68(3);\\
				B$_{12}$;B$_{13}$;B$_{23}$; &-0.01(2);0;0 &-0.02(2);0;0;  & -0.02(2);0;0; & -0.02(2);0;0; &-0.04(2);0;0;\\
				Rh2 B$_{11}$;B$_{22}$;B$_{33}$;&0.25(2);0.34(2);0.40(1); &0.32(1);0.41(2);0.43(1);  & 0.40(2);0.50(2);0.48(1); &0.49(2);0.59(2);0.45(2);  &0.46(2);0.59(2);0.56(2);\\
				B$_{12}$;B$_{13}$;B$_{23}$; &0.01(1);0.02(1);0.00(1) &0.00(1);0.02(1);0.01(1); & 0.001(1);0.03(1);0.01(1); & 0.00(2);0.02(2);0.01(2); &-0.00(2);0.04(2);0.01(2);\\
				Sn1 B$_{11}$;B$_{22}$;B$_{33}$; &0.47(2);0.42(2);0.44(2); &0.57(2);0.55(2);0.55(2);  &0.69(3);0.67(2);0.68(3);  & 0.79(4);0.85(3);0.78(3); &0.88(4);0.89(3);0.88(4);\\
				B$_{12}$;B$_{13}$;B$_{23}$;&-0.03(3);0.01(2);-0.02(3); &-0.02(3);0.02(2);-0.03(2);  & -0.03(3);0.03(2);0.00(3); & -0.02(4);0.05(3);-0.01(4); &-0.02(4);0.07(3);0.01(4);\\
				Sn2 B$_{11}$;B$_{22}$;B$_{33}$;&0.42(3);0.50(3);0.85(4); &0.51(3);0.60(3);0.86(4);  & 0.62(3);0.69(4);0.88(4); & 0.77(5);0.82(5);0.95(5); &0.74(5);0.81(5);1.02(5);\\
				B$_{12}$;B$_{13}$;B$_{23}$; &-0.15(2);-0.09(2);0.18(2); &-0.17(2);-0.09(2);0.16(2);  &-0.17(2);-0.06(2);0.12(2);  &-0.17(2);-0.05(3);0.10(3);  &-0.26(2);-0.06(3);0.14(3);\\
				Sn3 B$_{11}$;B$_{22}$;B$_{33}$;&0.44(3);0.45(3);0.91(5); &0.51(3);0.56(3);0.92(4);  & 0.62(3);0.66(3);0.93(4); & 0.76(4);0.83(4);1.00(5); &0.73(5);0.79(5);1.09(6);\\
				B$_{12}$;B$_{13}$;B$_{23}$; &-0.15(2);-0.03(2);-0.08(2)& -0.17(2);0.04(2);-0.08(2); &-0.18(2);0.04(2);-0.07(2)  & -0.17(2);0.03(3);-0.05(3); &-0.24(2);0.04(3);-0.08(3);\\
				Sn4 B$_{11}$;B$_{22}$;B$_{33}$;&0.44(1);0.44(1);0.60(1); & 0.56(1);0.55(1);0.68(1); & 0.69(2);0.68(2);0.77(1); &0.83(2);0.83(2);0.77(2);  &0.85(2);0.84(2);0.96(2);\\
				B$_{12}$;B$_{13}$;B$_{23}$; &0.151(9);-0.163(9);-0.153(9)&0.187(9);-0.199(9);-0.192(9);  &0.20(1);-0.21(1);-0.20(1);  &0.20(2);-0.21(2);-0.21(2);  &0.33(2);-0.34(2);-0.33(2);\\
				Sn5a B$_{11}$;B$_{22}$;B$_{33}$;&0.43(3);0.53(3);0.59(3); &0.51(3);0.58(3);0.64(3); & 0.62(3);0.67(3);0.72(3); & 0.73(4);0.72(4);0.70(4); &0.73(5);0.81(4);0.8485);\\
				B$_{12}$;B$_{13}$;B$_{23}$; &0.01(2);-0.11(3);0.02(2)& -0.01(2);-0.12(2);0.02(2);  &-0.01(2);-0.12(3);0.02(2);  & -0.02(3);-0.14(4);0.02(3); &-0.05(3);-0.22(4);0.05(3);\\
				Sn5b B$_{11}$;B$_{22}$;B$_{33}$;&0.65(3);0.48(4);0.54(3); &0.71(3);0.59(4);0.60(3);  & 0.80(3);0.69(4);0.70(4); & 0.85(4);0.79(5);0.69(5); &1.01(4);0.78(5);0.85(5);\\
				B$_{12}$;B$_{13}$;B$_{23}$;&-0.05(2);0.07(2);-0.11(3) &-0.06(2);0.06(2);-0.14(3)  & -0.05(3);0.05(3);-0.14(3); & -0.05(3);0.02(3);-0.16(4); &-0.04(3);0.05(3);-0.22(4);\\
				Sn6a B$_{11}$;B$_{22}$;B$_{33}$;&0.84(4);0.52(3);0.57(3); &0.81(4);0.60(3);0.63(3);  &0.83(4);0.69(4);0.71(4);  & 0.86(5);0.80(5);0.67(4); &1.02(5);0.82(5);0.83(5);\\
				B$_{12}$;B$_{13}$;B$_{23}$; &-0.16(2);-0.06(2);0.13(3); &-0.13(2);-0.05(2);0.14(3);  & -0.10(2);-0.03(2);0.15(3) &-0.08(3);-0.02(3);0.15(3)  &-0.10(3);-0.02(3);0.23(4);\\	
				Sn6b B$_{11}$;B$_{22}$;B$_{33}$;&0.78(3);0.41(3);0.58(3); & 0.84(3);0.50(3);0.64(3); &0.92(3);0.63(3);0.73(3);  &0.97(4);0.75(4);0.73(4);  &1.09(5);0.71(4);0.87(4);\\
				B$_{12}$;B$_{13}$;B$_{23}$; &0.07(2);0.10(2);0.12(2); &0.08(2);0.10(2);0.15(2);  &0.06(2);0.09(2);0.15(2);  &0.04(3);0.07(3);0.16(3)  &0.07(3);0.11(3);0.23(3);\\	
					\hline
		\end{tabular}
	\end{table} 
\end{landscape}

\clearpage

\begin{table*}
	\caption{Interatomic distances in the crystal structures of Sc$_\mathrm{5}$Rh$_\mathrm{6}$Sn$_\mathrm{18}$ assuming the Tb$_\mathrm{5}$Rh$_\mathrm{6}$Sn$_\mathrm{18}$ prototype and the Sc$_\mathrm{5}$Ir$_\mathrm{6}$Sn$_\mathrm{18}$ type (both SG \textit{I}4$_\mathrm{1}/acd$) as well as for a fictitious structure (SG \textit{I}4$_\mathrm{1}cd$). The numbers of atoms for the structure of Sc$_\mathrm{5}$Ir$_\mathrm{6}$Sn$_\mathrm{18}$ type are given in parentheses in accordance with Table 4. The atomic numbers with primes are from the split of the positions appearing due to \textit{group-subgroup} transformation of SG \textit{I}4$_\mathrm{1}/acd$ into \textit{I}4$_\mathrm{1}cd$.}
	\label{tbl:dist}
	\begin{tabularx}{\textwidth}{Xl@{\hspace{0.3cm}}Xl@{\hspace{0.3cm}}XXX}
		Atom&&Tb$_\mathrm{5}$Rh$_\mathrm{6}$Sn$_\mathrm{18}$ type&&Sc$_\mathrm{5}$Ir$_\mathrm{6}$Sn$_\mathrm{18}$ type&&Fictitious structure\\
		\hline
		\multirow{2}{\columnwidth}{Sc1}&-12Sn&3.345(1) - 3.363(1)&-24Sn&3.21(2) - 3.34(3)&-24Sn&3.212 - 3.524\\
		&-6Rh&3.486(1) - 3.491(1)&-6Rh&3.38(1) - 3.59(1)&-6Rh&3.383 - 3.575\\
		Sc1'&&&&&-24Sn&3.212 - 3.524\\
		\hline
		\multirow{2}{\columnwidth}{Sc2}&-3Rh&2.951(3) - 2.963(3)&-3Rh&2.950(2) - 2.961(2)&-3Rh&2.948 - 2.961\\
		&-10Sn&2.987(3) - 3.260(4)&-17Sn&2.995(2) - 3.360(2)&-17Sn&2.997 - 3.362\\
		Sc2'&&&&&-3Rh&2.951 - 2.965\\
		\hline
		\multirow{2}{\columnwidth}{Rh1}&-6Sn&2.618(1) - 2.686(1)&-10Sn&2.602(2) - 2.687(1)&-10Sn&2.603 - 2.688\\
		&-3Sc&2.959(3) - 3.486(1)&-8Sc&2.958(2) - 3.588(1)&-8Sc&2.959 - 3.588\\
		Rh1'&&&&&-10Sn&2.603 - 2.688\\
		\hline
		\multirow{2}{\columnwidth}{Rh2}&-6Sn&2.616(2) - 2.690(1)&-10Sn&2.608(2) - 2.691(1)&-10Sn&2.610 - 2.693\\
		&-3Sc&2.951(3) - 3.491(1)&-6Sc&2.950(2) - 3.57(3)&-6Sc&2.951 - 3.575\\
		Rh2'&&&&&-10Sn&2.610 - 2.693\\
		\hline
		\multirow{2}{\columnwidth}{Sn1}&-2Sc&2.987(3)&-2Sc&2.995(2)&-2Sc&2.997\\
		&-14Sn&3.237(2) - 4.142(2)&-24Sn&3.099(2) - 4.109(2)&-24Sn&3.101 - 4.109\\
		Sn1'&&&&&-2Sc&2.997\\
		\hline
		\multirow{3}{\columnwidth}{Sn2}&-2Rh&2.623(2)&-2Rh&2.610(1) - 2.647(1)&-2Rh&2.611 - 2.647\\
		&-9Sn&2.868(2) - 3.815(1)&-16Sn&2.865(2) - 4.109(2)&-16Sn&2.866 - 4.109\\
		&-3Sc&3.207(3)-3.353(1)&-3Sc&3.125(2) - 3.40(3)&-3Sc&3.126 - 3.406\\
		Sn2'&&&&&-2Rh&2.611 - 2.648\\
		\hline
		\multirow{3}{\columnwidth}{Sn3}&-2Rh&2.625(2)&-2Rh&2.623(1) - 2.642(1)&-2Rh&2.624 - 2.644\\
		&-9Sn&2.868(2) - 3.627(2)&-16Sn&2.865(2) - 3.845(2)&-16Sn&2.866 - 3.844\\
		&-3Sc&3.213(3) - 3.363(1)&-6Sc&3.118(2) - 3.45(3)&-6Sc&3.119 - 3.456\\
		Sn3'&&&&&-2Rh&2.624 - 2.644\\
		\hline
		\multirow{3}{\columnwidth}{Sn4}&-3Rh&2.686(1) - 2.690(1)&-3Rh&2.687(1) - 2.691(1)&-3Rh&2.688 - 2.691\\
		&-3Sc&3.120(3) - 3.133(3)&-3Sc&3.118(2) - 3.130(2)&-3Sc&3.120 - 3.131\\
		&-10Sn&3.355(1) - 3.654(2)&-16Sn&3.355(1) - 3.735(1)&-16Sn&3.356 - 3.735\\
		Sn4'&&&&&-3Rh&2.688 - 2.693\\
		\hline
		\multirow{3}{\columnwidth}{Sn5a}&-2Rh&2.618(1) - 2.624(2)&-2Rh&2.640(1) - 2.644(1)&-2Rh&2.641\\
		&-9Sn&2.894(2) - 4.142(2)&-13Sn&2.814(2) - 3.637(2)&-13Sn&2.814 - 3.638\\
		&-3Sc&3.250(4) - 3.345(1)&-6Sc&3.29(2) - 3.52(2)&-6Sc&3.296 - 3.524\\
		Sn5a'&&&&&-2Rh&2.641\\
		\hline
		\multirow{3}{\columnwidth}{Sn5b}&&&-2Rh&2.602(2) - 2.608(2)&-2Rh&2.603 - 2.610\\
		&&&-15Sn&2.915(2) - 4.080(2)&-15Sn&2.914 - 4.081\\
		&&&-6Sc&3.186(2) - 3.43(2)&-6Sc&3.188 - 3.431\\
		Sn5b'&&&&&-2Rh&2.603 - 2.606\\
		\hline
		\multirow{3}{\columnwidth}{Sn6a}&-2Rh&2.616(2) - 2.631(1)&-2Rh&2.617(1) - 2.665(1)&-2Rh&2.617 - 2.666\\
		&-9Sn&2.863(2) - 3.740(2)&-16Sn&2.862(2) - 3.927(2)&-16Sn&2.861 - 3.929\\
		&-3Sc&3.173(3) - 3.363(1)&-6Sc&3.078(2) - 3.48(2)&-6Sc&3.078 - 3.481\\
		Sn6a'&&&&&-2Rh&2.622 - 2.666\\
		\hline
		\multirow{3}{\columnwidth}{Sn6b}&&&-2Rh&2.614(1) - 2.625(1)&-2Rh&2.615 - 2.624\\
		&&&-16Sn&2.862(2) - 4.039(2)&-16Sn&2.861 - 4.040\\
		&&&-6Sc&3.177(2) - 3.45(2)&-6Sc&3.179 - 3.448\\
		Sn6b'&&&&&-2Rh&2.615 - 2.629\\
		\hline
	\end{tabularx}
\end{table*} 

\clearpage

\begin{table}
	\caption{Comparison of effective charges computed for the ideal and split models}
	\centering
	\label{tbl:effchargecompar}
	\begin{tabular}{c@{\hspace{1cm}}c@{\hspace{1cm}}c@{\hspace{1cm}}c}
		Wyckoff position (Ideal)& Effective Charge & Wyckoff Position (Split) & Effective Charge \\
		\hline
		Sc1 					& +1.32			 		 & Sc1  & +1.30   \\
		\hline
		\multirow{2}{*}{Sc2} 	& \multirow{2}{*}{+1.30} & Sc2  & +1.28  \\
							 	&						 & Sc2' & +1.29  \\
		\hline
		\multirow{2}{*}{Rh1} 	& \multirow{2}{*}{-1.18} & Rh1  & -1.15  \\
								&						 & Rh1' & -1.17  \\
		\hline
		\multirow{2}{*}{Rh2} 	& \multirow{2}{*}{-1.17} & Rh2  & -1.15  \\
								&						 & Rh2' & -1.15  \\
		\hline
		Sn1 					& -0.53					 & Sn1  & -0.53  \\
		\hline
		Sn2 					& +0.05					 & Sn2  & +0.05  \\
		\hline
		Sn3 					& +0.14					 & Sn3  & +0.11  \\
		\hline
		\multirow{2}{*}{Sn4} 	& \multirow{2}{*}{+0.18} & Sn4  & +0.17   \\
								&						 & Sn4' & +0.16   \\
		\hline
		\multirow{2}{*}{Sn5} 	& \multirow{2}{*}{+0.06} & Sn5  & +0.01   \\
								&						 & Sn5' & +0.08   \\
		\hline
		\multirow{2}{*}{Sn6} 	& \multirow{2}{*}{+0.06} & Sn6  & +0.11   \\
								&						 & Sn6' & +0.08   \\
		\hline
	\end{tabular}
\end{table} 

\twocolumn
\subsection{Figures}

\begin{figure}[h]
	\includegraphics[width=\linewidth]{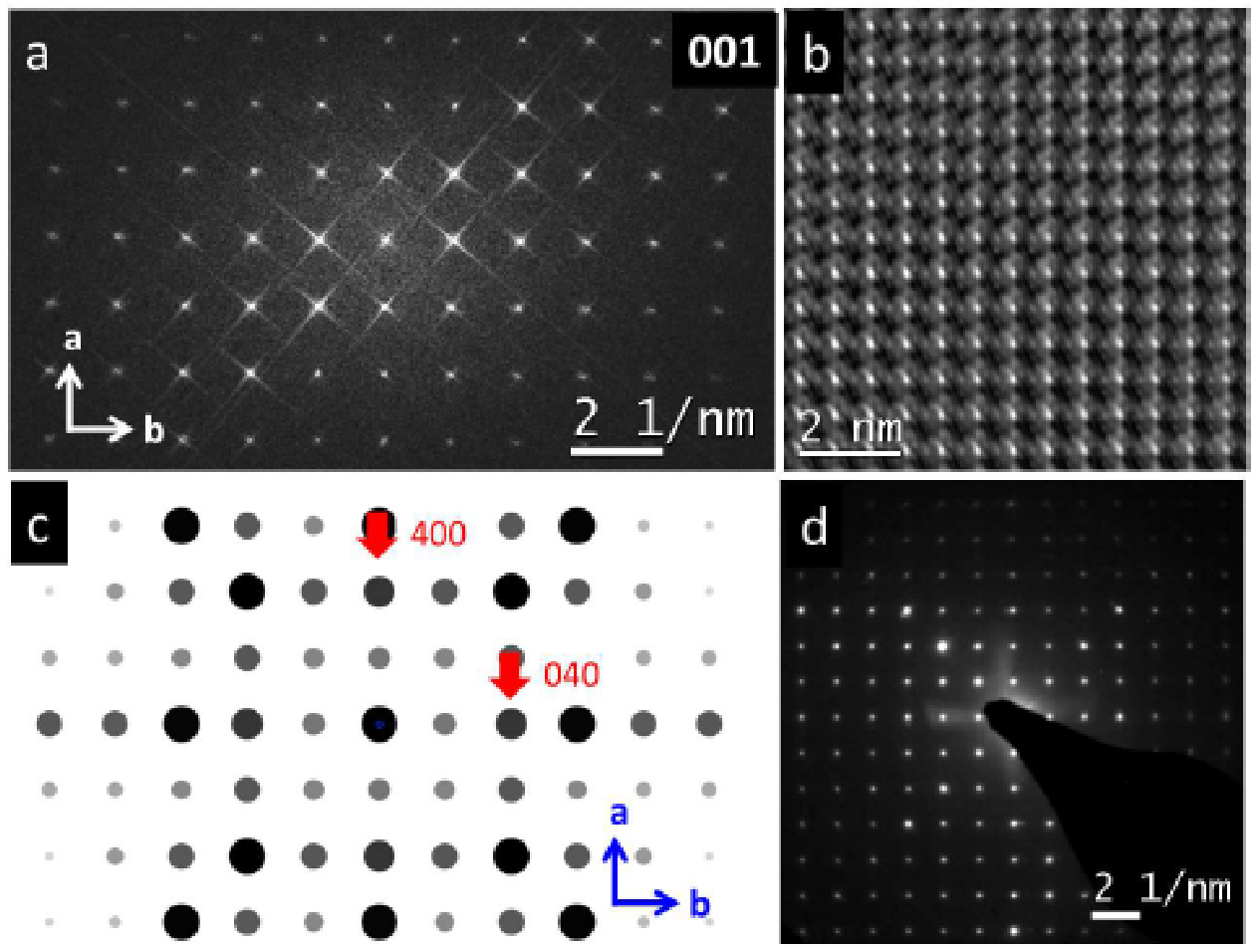}
	\caption{: TEM of the [0\,0\,1] zone in Sc$_5$Rh$_6$Sn$_{18}$. (a) Fourier transform of the high-resolution image. No superstructure reflections were found in this zone. (b) Corresponding Fourier filtered high-resolution image. (c) Simulated diffraction pattern of the [0\,0\,1] zone. (d) Electron diffraction pattern of the [0\,0\,1] zone. Intensities appear equalized due to dynamic scattering compared to the kinematic simulation in (c).}
	\label{fig:TEM001}
\end{figure}

\begin{figure}
	\includegraphics[width=\linewidth]{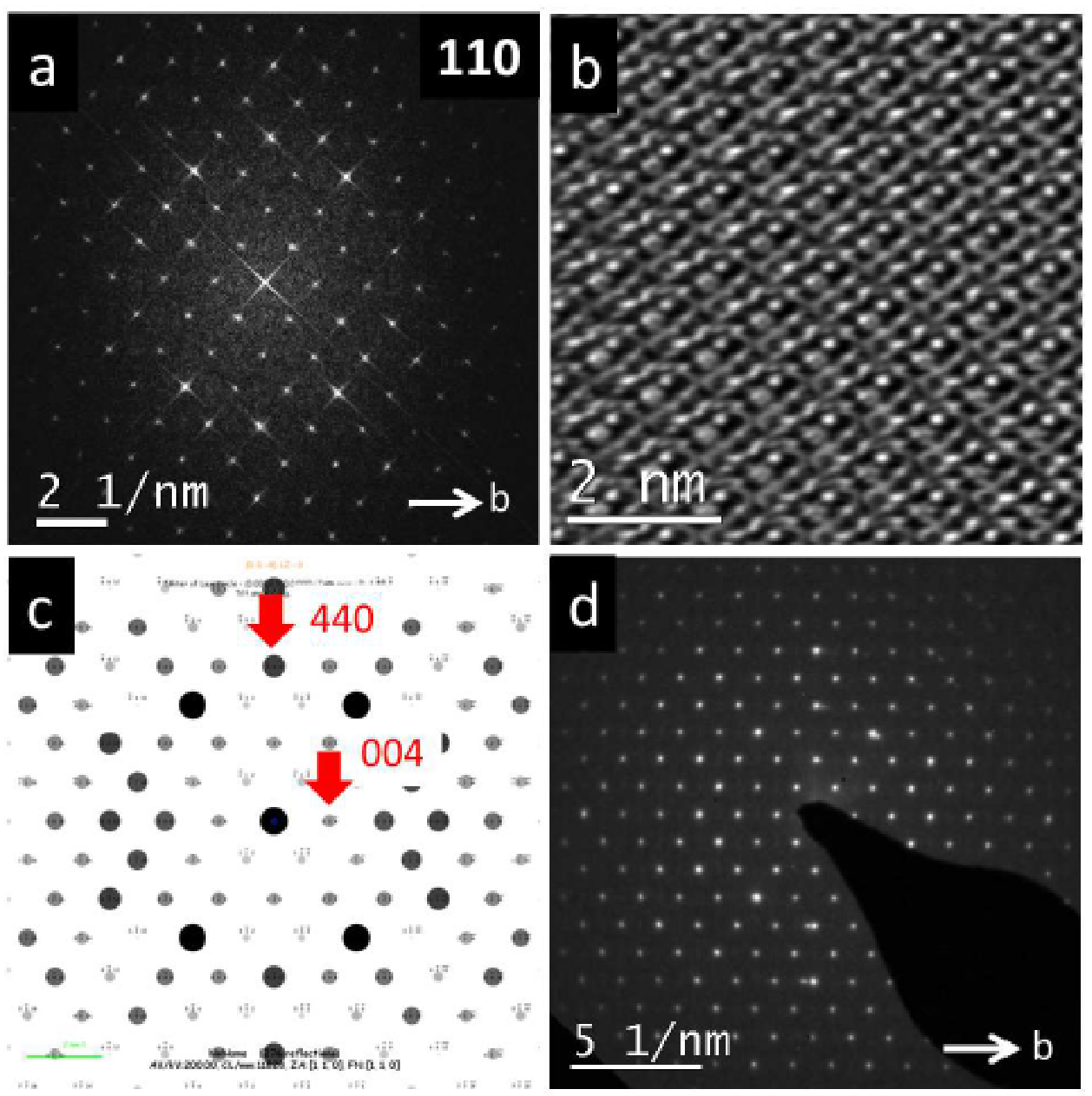}
	\caption{: TEM of the [1\,1\,0] zone in Sc$_5$Rh$_6$Sn$_{18}$. (a) Fourier transform of the high-resolution image. No superstructure reflections were found in this zone. (b) Corresponding Fourier filtered high-resolution image. (c) Simulated diffraction pattern of the [1\,1\,0] zone. (d) Electron diffraction pattern of the [1\,1\,0] zone. Intensities appear equalized due to dynamic scattering compared to the kinematic simulation in (c).}
	\label{fig:TEM110}
\end{figure}

\begin{figure}
	\includegraphics[width=0.8\linewidth]{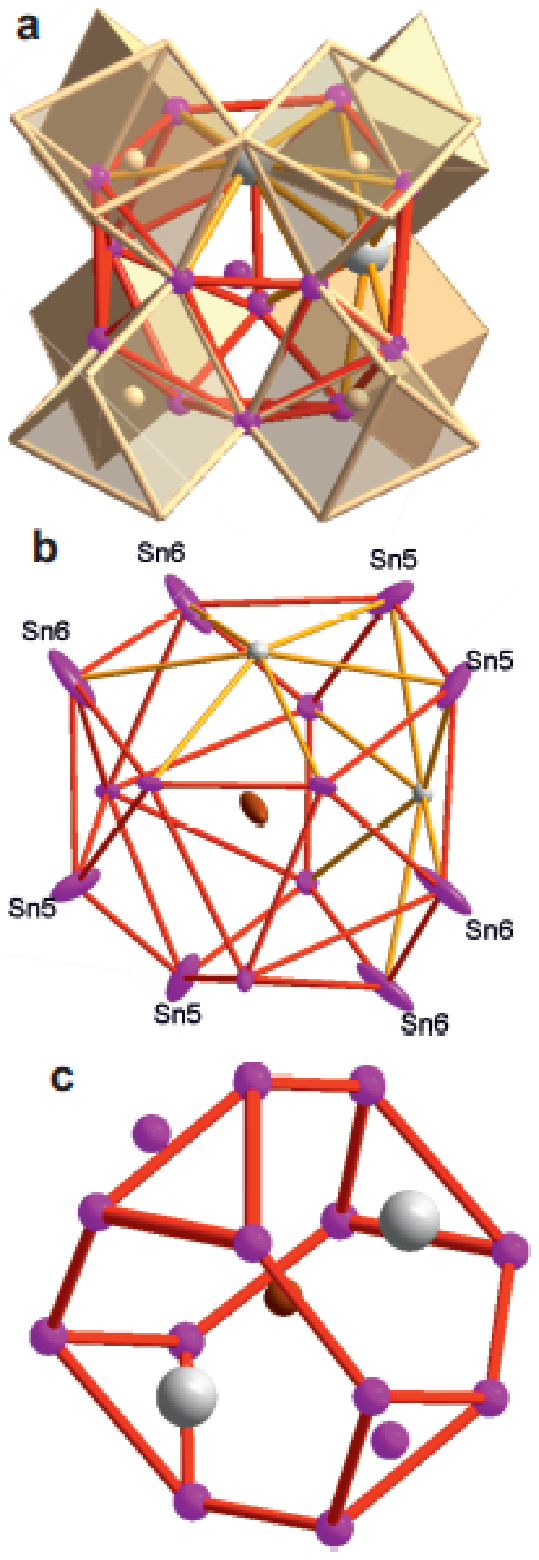}
	\caption{First coordination sphere of the Sn1-atom. a: The Sc$_2$Sn$_{14}$ polyhedron incorporated in trigonal prisms. b: Illustration (ellipsoid probability: 90\,\%) of the anisotropic displacement parameters (see Table \ref{tbl:bvalues}) of all atoms forming the vertices of the polyhedron. c: Sn1Sc2$_2$Sn$_{14}$ Frank-Kasper polyhedron}
	\label{fig:Anis}
\end{figure}

\clearpage
\begin{landscape}
\begin{figure}
	\includegraphics[width=\linewidth]{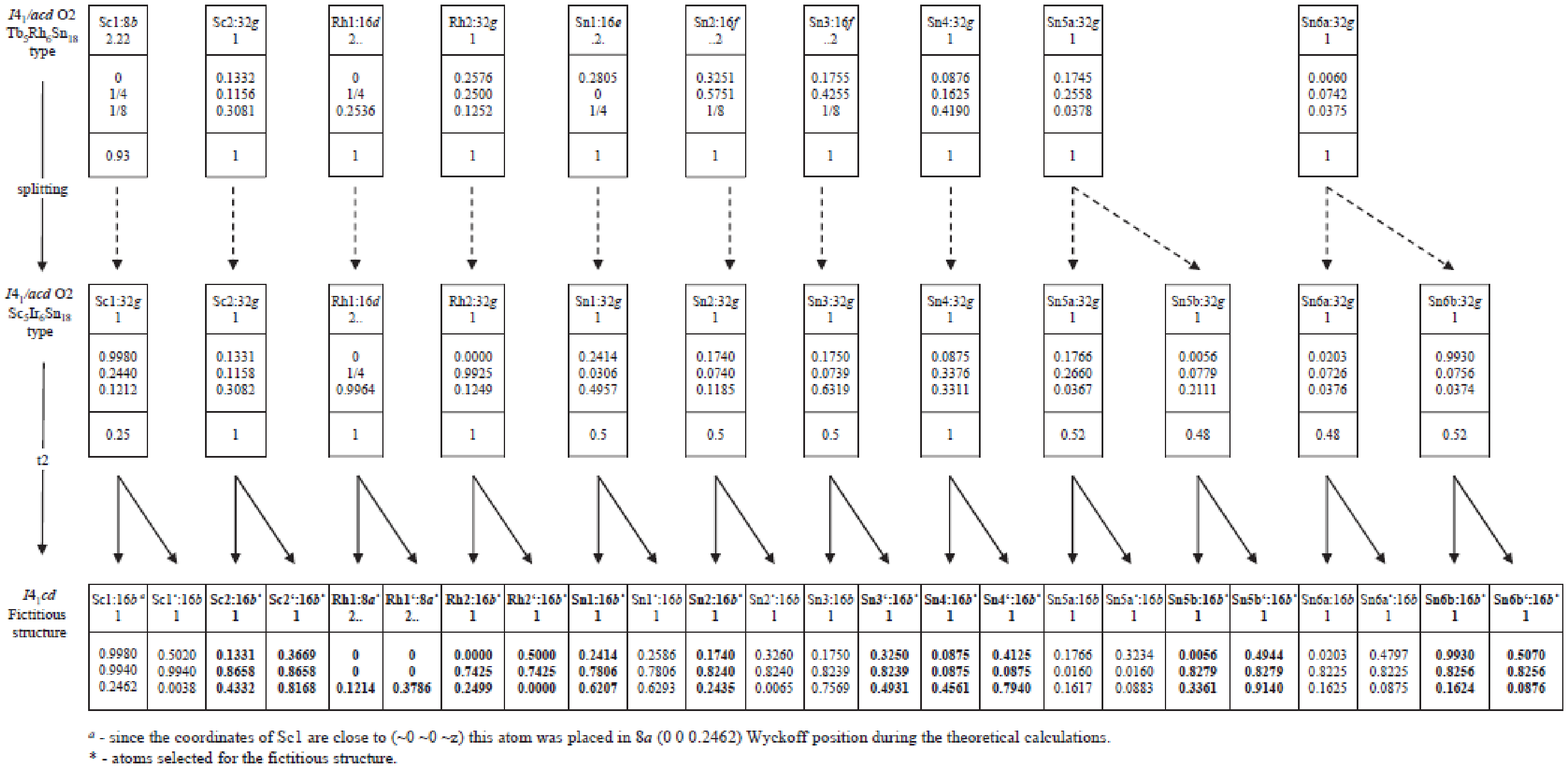}
	\caption{Comparison of the structural models for the Sc$_\mathrm{5}$Rh$_\mathrm{6}$Sn$_\mathrm{18}$ stannide given as the B\"arnighausen scheme\cite{baerninghausen1980,Mueller2004,Mueller2013} including the crystallographic sites, site symmetry, atomic coordinates and occupational parameters \textit{G} (not given for the fictitious structure because \textit{G} = 1). Dashed lines indicate a splitting of atomic positions within the same space group. The fictitious structure (SG \textit{I}4$_\mathrm{1}cd$) was obtained for the chemical bonding analysis from the Sc$_\mathrm{5}$Ir$_\mathrm{6}$Sn$_\mathrm{18}$ type by a \textit{group-subgroup} scheme (see text for more details).}
	\label{fig:Baernighausen}
\end{figure}
\end{landscape}

\clearpage

\balance

\providecommand*{\mcitethebibliography}{\thebibliography}
\csname @ifundefined\endcsname{endmcitethebibliography}
{\let\endmcitethebibliography\endthebibliography}{}

\end{document}